\documentclass[twocolumn]{openjournal}

\makeatletter
\def\doauthor#1#2#3{%
  \ignorespaces#1\unskip
  \begingroup
    #3%
  \@if@empty{#2}
    {\endgroup{}{}}
    {\endgroup{\textsuperscript{,*}}{}%
      \begingroup
        \def\@thefnmark{*}%
        \frontmatter@footnotetext{#2}%
      \endgroup}%
  \@listcomma\space\@listand
}
\makeatother

\usepackage{etoolbox}

\makeatletter

\renewcommand{\apjsecfont}{\normalsize\bfseries}
\patchcmd{\section}
  {9pt plus 2pt minus 1pt}
  {14pt plus 2pt minus 1pt}
  {}{}
\patchcmd{\section}
  {4pt}
  {6pt}
  {}{}

\makeatother

\usepackage{xcolor}
\usepackage{textgreek}
\usepackage[utf8]{inputenc}
\usepackage[english]{babel}

\usepackage{relsize}
\usepackage{pifont}
\usepackage{ragged2e}

\usepackage{hyperref}
\hypersetup{
    unicode, 
    colorlinks=true,
    linkcolor=linkcolor,
    citecolor=linkcolor,
    filecolor=linkcolor,
    urlcolor=linkcolor,
}
\usepackage{color,colortbl}
\definecolor{linkcolor}{RGB}{29, 132, 181} 

\usepackage{tensind}
\tensordelimiter{?}
\DeclareGraphicsExtensions{.bmp,.png,.jpg,.pdf}
\usepackage{verbatim}
\usepackage[normalem]{ulem}
\usepackage{orcidlink}
\usepackage{soul}

\usepackage{placeins}

\newcommand{\Msun}[1]{\mathrm{M}_\odot}
\newcommand{\Mstar}[1]{\mathrm{M}_\star}
\newcommand{\ntracer}[1]{n_\mathrm{tracer}}
\newcommand{\Rfil}[1]{R_\mathrm{fil}}

\graphicspath{ {./figs/} }

\begin{document}
\title[Gas line-mass: unifying cosmic filaments]{\textbf{Linear gas mass density: a unifying descriptor of Cosmic Filaments}}


\author{Daniela Galárraga-Espinosa\orcidlink{0000-0002-8808-803X}}
\email{daniela.galarraga@ipmu.jp}
\affiliation{Kavli IPMU (WPI), UTIAS, The University of Tokyo,
Kashiwa, Chiba 277-8583, Japan}
\affiliation{Center for Data-Driven Discovery, Kavli IPMU (WPI), UTIAS,
The University of Tokyo, Kashiwa, Chiba 277-8583, Japan}

\author{Daisuke Nagai\orcidlink{0000-0002-6766-5942}}
\affiliation{Department of Physics, Yale University,
New Haven, CT 06520, USA}
\affiliation{Department of Astronomy, Yale University,
New Haven, CT 06520, USA}

\author{Isabel Medlock\orcidlink{0009-0000-7415-8239}}
\affiliation{Department of Astronomy, Yale University,
New Haven, CT 06520, USA}

\author{Mathieu Langer\orcidlink{0000-0002-9088-2718}}
\affiliation{LUX, Observatoire de Paris, Université PSL,
Sorbonne Université, CNRS, 75014 Paris, France}

\author{Nabila Aghanim\orcidlink{0000-0002-6688-8992}}
\affiliation{Université Paris-Saclay, CNRS,
Institut d'Astrophysique Spatiale, 91405 Orsay, France}

\author{Kentaro Nagamine\orcidlink{0000-0001-7457-8487}}
\affiliation{Theoretical Astrophysics, Department of Earth \& Space Science,
Graduate School of Science, The University of Osaka,
1-1 Machikaneyama, Toyonaka, Osaka 560-0043, Japan}
\affiliation{Theoretical Joint Research, Forefront Research Center,
Graduate School of Science, The University of Osaka,
1-1 Machikaneyama, Toyonaka, Osaka 560-0043, Japan}
\affiliation{Kavli IPMU (WPI), UTIAS, The University of Tokyo,
Kashiwa, Chiba 277-8583, Japan}
\affiliation{Department of Physics \& Astronomy,
University of Nevada, Las Vegas,
4505 S. Maryland Pkwy, Las Vegas, NV 89154-4002, USA}
\affiliation{Nevada Center for Astrophysics,
University of Nevada, Las Vegas,
4505 S. Maryland Pkwy, Las Vegas, NV 89154-4002, USA}


\begin{abstract}
    Cosmic filaments connect the densest nodes of the large-scale web and are thought to harbour a substantial fraction of the low-redshift baryon budget, yet a physically motivated framework for organising their diverse gas properties has been lacking. Here we demonstrate that the gaseous linear mass density $\Lambda \equiv M_{\rm gas}/L_{\rm fil}$, measured within the entropy radius $r_K$ --the location of the maximum in the radial gas entropy profile-- plays this role, analogous to mass in halo physics. Using IllustrisTNG300-1 at $z = 0$, we characterise the structural, thermodynamic, and chemical properties of $5637$ filaments spanning more than two decades in $\Lambda$. Filaments form a continuous thermodynamic sequence with $\Lambda$, ranging from long, thin, cool, metal-poor structures to shorter, wider, hotter, and more enriched ones, with core temperatures spanning nearly two orders of magnitude across the full WHIM window. 
    The ratio of mass- to volume-weighted temperature reveals a clear change in how heat is distributed within filaments: low-$\Lambda$ filaments have their hotter gas preferentially associated with denser material, whereas in high-$\Lambda$ filaments the hotter gas is preferentially diffuse. Well-defined scaling relations connect $\Lambda$ to temperature, electron density, pressure, entropy, and metallicity, with the  $T$--$\Lambda$ scaling qualitatively consistent with gravitational heating.
    We present the first measurement of the gaseous linear mass density function, revealing a broad distribution with a flat low-$\Lambda$ regime and a steep exponential decline at high $\Lambda$, analogous to the halo mass function. The integrated Compton-$y$ parameter $Y^{\rm int}$ scales systematically with $\Lambda$, providing a novel tSZ proxy accessible to current surveys. Finally, galaxy richness per unit length correlates tightly with $\Lambda$ (Pearson $r = 0.80$), enabling $\Lambda$-based classification of filaments from spectroscopic surveys alone and guiding targeted searches for the missing baryons in the cosmic web.
\end{abstract}



\maketitle


\section{Introduction}
\label{Sect:Introduction}

The matter distribution in the Universe forms a complex network of knots, filaments, sheets, and voids collectively known as the cosmic web \citep{White1987, Bond1996}. This structure arises naturally from the anisotropic gravitational collapse of primordial density fluctuations, with matter flowing from underdense regions towards sheets and filaments and ultimately into the massive haloes located at their intersections \citep{Zeldovich1970}. Cosmic filaments occupy a central position in the assembly of large-scale structure: they connect the densest environments of the Universe, channel matter towards haloes and galaxies, and contain a substantial fraction of the matter outside collapsed structures \citep[e.g.,][]{Cautun2014}.

The baryonic component of the filaments is particularly important in this context. Cosmological hydrodynamical simulations predict that a significant fraction of low-redshift baryons resides in the warm-hot intergalactic medium (WHIM), with temperatures of approximately $10^5$--$10^7$ K \citep[e.g.,][]{CenOstriker2006,Dave2001_massing_baryons, Wang2026_whim}. A large fraction of this gas is associated with the filamentary cosmic web and is difficult to detect directly because of its low density and temperature. Filaments have consequently become important targets for searches for the so-called missing baryons through, among others, X-ray emission and absorption and the thermal Sunyaev--Zel'dovich effect \citep[e.g.,][]{Nicastro2018Nature,DeGraaff2019,Tanimura2019}. Statistical detections of their X-ray emission have recently been obtained with ROSAT and eROSITA over large filament samples \citep{Tanimura2020_Rosat,Tanimura2022_eRosita,Zhang2024erosita_fils}. At the same time, observations of individual structures, including inter-cluster filaments such as A399-A401 \citep{Bonjean2018} and filaments in the Shapley supercluster \citep{Aghanim2024_shapley,Migkas2025}, are beginning to provide direct constraints on the properties of specific filament environments.
Yet despite this progress, the physical properties of filament gas remain poorly characterised both observationally and theoretically, in part because the filament population spans a wide and poorly understood range of thermodynamic states. Numerical studies are therefore essential both for establishing these properties and for connecting them to observable tracers.

In \citet{GalarragaEspinosa2020}, we showed that filaments selected according to their length can trace statistically distinct environments: short filaments are typically denser, wider, and more strongly connected to massive structures, whereas long filaments are thinner and preferentially populate less dense regions of the cosmic web.
The diversity of the filament population has also been characterised through other structural properties, like filament thickness or diameter \citep[e.g.,][]{Colberg2005,Cautun2014,GaneshaiahVeena2018_CosmicBalletI}. Filament diversity is also reflected in their baryonic properties. Indeed, several studies in numerical simulations have shown that the gas content, temperature, and pressure of the filaments depend on the filament length \citep{GalarragaEspinosa2021,GalarragaEspinosa2022}, thickness, and prominence of the structure \citep[e.g.,][]{GhellerVazza2015, Zhu2021_filaments_evo_z, Tuominen2021}.

Much of our current knowledge of filament properties, however, has been obtained from average or stacked measurements. Stacking is particularly powerful for these diffuse structures, both numerically and observationally, because it increases the signal-to-noise ratio and reveals their average radial structure. At the same time, it necessarily compresses the intrinsic diversity of the filament population. Relations between filament properties are then often inferred by dividing the population according to a predefined quantity, such as length or radius, rather than by identifying the physical variable that most fundamentally organises the population. What is missing is a single physically motivated parameter that organises the full thermodynamic diversity of the filament population --as halo mass does for galaxy clusters-- and connects it directly to observational tracers accessible to current surveys. \\

In this work, we show that the gaseous linear mass density $\Lambda \equiv M_{\rm gas}/L_{\rm fil}$, measured within a physically motivated thermodynamic boundary, provides this unifying parameter.
The linear mass density (hereafter also called `line-mass', for brevity) is a fundamental quantity in the dynamics of self-gravitating cylindrical systems. In idealised models of isolated and cylindrical filaments, $\Lambda$ controls the gravitational potential and is directly related to the filaments' equilibrium and stability properties \citep[e.g.,][]{Stodolkiewicz1963,Ostriker1964,Inutsuka1992}.
More recent theoretical and numerical studies have extended this framework to intergalactic streams and cosmic-web filaments, showing connections between line-mass, temperature, gravitational confinement, and filament dynamics \citep[e.g.,][]{Aung2019, Lu2024_fils}. Although cosmological filaments are neither isolated nor perfectly cylindrical equilibrium systems, these results motivate $\Lambda$ as a physically meaningful quantity with which to organise their internal properties.\\

We perform this study using the IllustrisTNG hydrodynamical simulation and we focus on the low-redshift filament population ($z=0$). We use a filament catalogue detected from the galaxy distribution rather than from the underlying dark-matter field in order to mirror the situation faced by observations, where the dark-matter density field is not directly accessible, and filaments must instead be reconstructed from galaxies. Our methodology is thus designed with applications to present and upcoming large spectroscopic surveys in mind, including Subaru PFS, DESI, and SDSS.

The paper is organised as follows. Sect.~\ref{Sect:DataMethods} describes the simulation, filament catalogue, and gas property measurements. Sect.~\ref{Sect:LambdaAndRadius} introduces the entropy radius $r_K$, defines the gaseous line-mass $\Lambda$, and presents the first measurement of the filament line-mass function. In Sect.~\ref{Sect:FilProperties} we characterise the structural, thermodynamic, and chemical properties of filaments as a function of $\Lambda$, we derive the corresponding scaling relations, and demonstrate a systematic transition in the internal thermal structure of filament gas. Section~\ref{Sect:Observables} presents two complementary observational routes to $\Lambda$: a galaxy richness proxy accessible from spectroscopic surveys (Sect.~\ref{SubSect:GalaxyProxy}) and the integrated Compton-$y$ parameter $Y^{\rm int}$ as a tSZ probe of filament thermodynamics (Sect.~\ref{SubSect:Yint}).

\section{Data and methods}\label{Sect:DataMethods}

\subsection{IllustrisTNG300-1 simulations}

The IllustrisTNG project is a suite of large cosmological magneto-hydrodynamical simulations of galaxy formation and evolution \citep{Springel2018_TNG,Pillepich2018_TNGgalaxies,Nelson2018,Marinacci2018}. The simulations are performed with the moving-mesh code AREPO \citep{Springel2010} and follow the coupled evolution of dark matter, gas, stars, and black holes. The TNG galaxy-formation model includes radiative gas cooling and heating, star formation and stellar evolution, chemical enrichment, stellar feedback, and the formation, growth, and feedback of supermassive black holes. The simulations adopt cosmological parameters consistent with the Planck 2015 results \citep{Planck2015Cosmo}.

In this work, we use TNG300-1, the highest-resolution realisation of the TNG300 volume. It follows a periodic cubic volume of side length 205 cMpc/h $\simeq 303$ cMpc, initially sampled by $2500^3$ DM particles and approximately the same number of gas cells. The DM particle mass is $m_\mathrm{DM} \simeq 5.9 \times10^7$ $\Msun{}$, while the target baryonic mass resolution is $m_\mathrm{gas} \simeq 1.1 \times 10^7$ $\Msun{}$. 
TNG300-1 is uniquely suited to the present studt as it combines a large simulated volume with sufficient resolution to study the gaseous properties of individual filaments across a broad range of very large-scale environments. Its volume provides a statistically large filament population, while its resolution allows their thermodynamic structure to be characterised down to sub-Mpc scales (see Sect.~\ref{SubSect:ResolutionCut}). For our analysis, we use the full gas-cell dataset of TNG300-1, together with the Subfind subhalo catalogues described below. This work focuses on the $z=0$ snapshot.

\subsection{\label{SubSect:GasProperties}Gas properties ($T$, $P_e$, $n_e$, $Z$, $K$)}

The gas properties studied in this work are derived for individual gas cells (\texttt{PartType0}) using the fields available in the IllustrisTNG snapshots. Details of the simulation fields entering the following definitions are provided in the IllustrisTNG data specifications\footnote{\url{https://www.tng-project.org/data/docs/specifications/}}. Unless stated otherwise, all quantities are converted to International System (SI) units before performing the calculations.

We estimate the gas temperature by assuming a perfect monoatomic gas,
\begin{equation}
T = (\gamma -1) \, \frac{u}{k_\mathrm{B}} \,\mu,
\end{equation}
where $\gamma=5/3$ is the adiabatic index, $u$ is the specific internal energy of the gas cell (\texttt{InternalEnergy}), $k_\mathrm{B}$ is the Boltzmann constant, and $\mu = [4/(1 + 3X_{\rm H} + 4X_{\rm H}x_e)] \times m_p$ is the mean molecular weight, with hydrogen mass fraction $X_{\rm H} = 0.76$, electron abundance $x_e$ (\texttt{ElectronAbundance}), and proton mass $m_\mathrm{p}$.
The electron number density is then $n_e =  x_e\,X_{\rm H}\,\rho_{\rm cell}/m_p$, where $\rho_{\rm cell}$ is the gas cell density (\texttt{Density}), and
$\mu_e \approx 2/(1+X_{\rm H})\simeq 1.14$ is the mean molecular weight per electron.
The corresponding electron pressure is then given by
\begin{equation}
P_e = k_\mathrm{B} \, n_e \, T.
\end{equation}

We additionally consider the entropy proxy
\begin{equation}\label{Eq:Entropy}
K = k_\mathrm{B} \, T \, n_e^{-2/3}.
\end{equation}
This quantity provides a useful tracer of the thermodynamic history of the filament gas, because it is conserved during adiabatic compression or expansion, but changes when the gas is heated or cooled, for example by shocks, radiative cooling, or feedback \citep[e.g.,][in the context of galaxy clusters]{Pratt2010}. Entropy 
therefore contains information on both the gravitational assembly of filaments and the additional physical processes affecting their gaseous component. 
We note that $K$ is an entropy proxy rather than the thermodynamic entropy itself. For an ideal monoatomic gas of fixed composition, $K$ is monotonically related to the entropy through $s \propto \ln K$.

Finally, the gas metallicity, $Z$, is taken directly from the \texttt{GFM\_Metallicity} field and corresponds to the mass fraction of elements heavier than helium in each gas cell. We express metallicities in solar units by adopting $Z_\odot=0.0127$ \citep{Asplund2009}.
Because metal enrichment depends sensitively on the adopted stellar evolution and feedback prescriptions, the metallicities reported here should be interpreted as predictions of the TNG model rather than as universal values for cosmic filaments.

Throughout this work, we report electron number density, electron pressure, and entropy in units of $\mathrm{cm^{-3}}$, $\mathrm{keV\,cm^{-3}}$, and $\mathrm{keV\,cm^2}$, respectively.

\subsection{Filament catalogue}\label{SubSect:filament_Cat}

We use the filament catalogue traced by the galaxy distribution, following \citet{GalarragaEspinosa2020}. Tracer galaxies are selected from the \textsc{Subfind} subhalo catalogue of the $z=0$ TNG300-1 snapshot, with a minimum stellar mass of $10^{9} \, \Msun{}$. The filamentary network was extracted with the filament finder DisPerSE \citep{Sousbie2011, Sousbie2011b} using a persistence threshold of $3\sigma$. This threshold was calibrated following the method of \citet{GalarragaEspinosa2024_MTNG}. The $3\sigma$ skeleton maximises the recovery of physically meaningful filaments whose end points correspond to the postions of galaxy clusters, while suppressing noise-driven topological features in the galaxy density field.
Tracing the filament network from the galaxy distribution rather than the dark matter field is a deliberate methodological choice. It mirrors the information available to observers and ensures that the filament catalogue is more directly comparable to those constructed from spectroscopic surveys such as SDSS, DESI, and Subaru PFS. The minimum stellar mass threshold of $10^9\,\Msun{}$ is chosen to roughly match the galaxy populations accessible to these surveys at low redshift.

Spurious structures are removed following the cleaning procedure described in \citet{GalarragaEspinosa2024_MTNG}. This step removes an excess of critical points and associated filaments located near the boundaries of the simulation box, which can arise despite the use of periodic boundary conditions. Previous tests showed that the structures removed by this procedure are numerical artefacts. We retain complete filaments and therefore do not activate the DisPerSE `breakdown' option. The resulting catalogue contains $10115$ filaments. These filaments span a broad range of lengths and environments, collectively tracing the full diversity of large-scale structure in the TNG300-1 volume from underdense voids to the outskirts of massive nodes. This provides the statistical basis for the line-mass analysis presented in Sect.~\ref{SubSect:LineMassFunction}.

To isolate the filament signal from that of massive nodes of the cosmic web, we further mask filament segments located in their vicinity. We discard segments lying within $3 \, R_{200c}$ of galaxy clusters, defined here as FoF haloes with $M_{200c}>5\times10^{13}$ $\Msun{} / h$, as well as segments located within 1~Mpc of topological maxima \citep[CPmax,][]{GalarragaEspinosa2020}. This masking removes some of the shortest structures entirely, including bridges inside neighbouring galaxy clusters. After applying this mask, $9019$ filaments remain from the initial sample of $10115$.

\subsection{Computing the profiles using all available gas cells}\label{SubSect:Method_Profiles}

For each filament, we compute radial profiles of the gas temperature, pressure, electron density, metallicity, and entropy using the full set of available gas cells. The profiles are measured as a function of the perpendicular distance to the filament spine, hereafer $r$, up to a maximum radius of $r=4$~Mpc. We adopt fixed linear radial bins of width 150~kpc. This choice is motivated by resolution considerations: we verified that the effective radii of gas cells in the vicinity of the filaments remain below approximately 115~kpc, ensuring that the adopted bin size does not probe scales smaller than the typical gas cell resolution.

For a given gas quantity $Q = T, P_e, n_e, Z, K$ (defined in Sect.~\ref{SubSect:GasProperties}), we compute its radial profile using volume-weighted averages of individual gas cells, following \cite{GalarragaEspinosa2021}:
\begin{equation}\label{Eq:Qprofile}
    Q(r_i) = \frac{\displaystyle
    \sum_{s = 1}^{N_\mathrm{seg}}
   \left( \sum_{j = 1}^{N(i)} Q^j \times V^j \right)_{s}
    }{
    \displaystyle \sum_{s = 1}^{N_\mathrm{seg}} \left( \sum_{j = 1}^{N(i)} V^j \right)_{s} 
    } \ \ ,
\end{equation}
where $Q^j$ and $V^j$ correspond respectively to the value and volume of the $j$-th gas cell, 
$N(i)$ is the number of cells inside the $i$-th cylindrical shell around the segment $s$, and $r_i$ indicates the centre of the radial bin.
Effectively, Eq.~\ref{Eq:Qprofile} gives the mean radial profile of $Q$ around the filament spine: for each radial bin, $Q$ is averaged over the gas cells contained in the corresponding cylindrical shells and combined over the full set of segments composing the filament.
We note that for a study of combined quantities (e.g., mean $T \times P$), a simple multiplication of individual profiles would be incorrect, since one would need to apply again Eq.~\ref{Eq:Qprofile} to the combination of variables at the cell level.\\

In the case of temperature, we also consider the mass-weighted average profile, $T_m(r)$. While the volume-weighted temperature, hereafter $T_v$, traces the temperature of the volume-filling gas, the mass-weighted temperature is dominated by the denser gas instead (i.e., by the filament regions that contain most of the baryonic mass). The mass-weighted average is computed as:
\begin{equation}\label{Eq:Tmprofile}
    T_m(r_i) = \frac{\displaystyle
    \sum_{s = 1}^{N_\mathrm{seg}}
   \left( \sum_{j = 1}^{N(i)} T^j \times M_\mathrm{gas}^j \right)_{s}
    }{
    \displaystyle \sum_{s = 1}^{N_\mathrm{seg}} \left( \sum_{j = 1}^{N(i)} M^j_\mathrm{gas} \right)_{s} 
    },
\end{equation}
where $M^j_\mathrm{gas}=  m_p \, \mu_e \, n_e^j \, V^j$ is the gas mass of the $j$-th gas cell, and $\mu_e = 1.14$ is the mean molecular weight per free electron.

Computing the gas property profiles using all available cells is computationally very expensive and, to our knowledge, this is the first time it is done for a sample of several thousands of filaments. The main difficulty comes from the sheer number of gas cells in high-resolution hydrodynamical simulations such as the one used in this work. Within the AREPO framework \citep{Springel2010}, gas cells are stored across multiple data chunks that do not correspond to a regular partition of the simulation volume. Since filaments are elongated structures extending over several Mpc, they can intersect many of these chunks, making it challenging to identify and access all the gas cells required to compute a complete radial profile.\\

In addition to the filament portions described above, we mask the gas cells located inside $M_{200c} > 5 \times 10^{13} \, \Msun~/h$ haloes (within $3 \times R_{200}$) and the nodes of the web (within 1~Mpc) before calculating the profile. This conservative mask ensures that we focus on filament gas without contamination by hotter, denser gas residing in and around massive haloes.\\

\begin{figure*}
	\includegraphics[width=1.\linewidth]{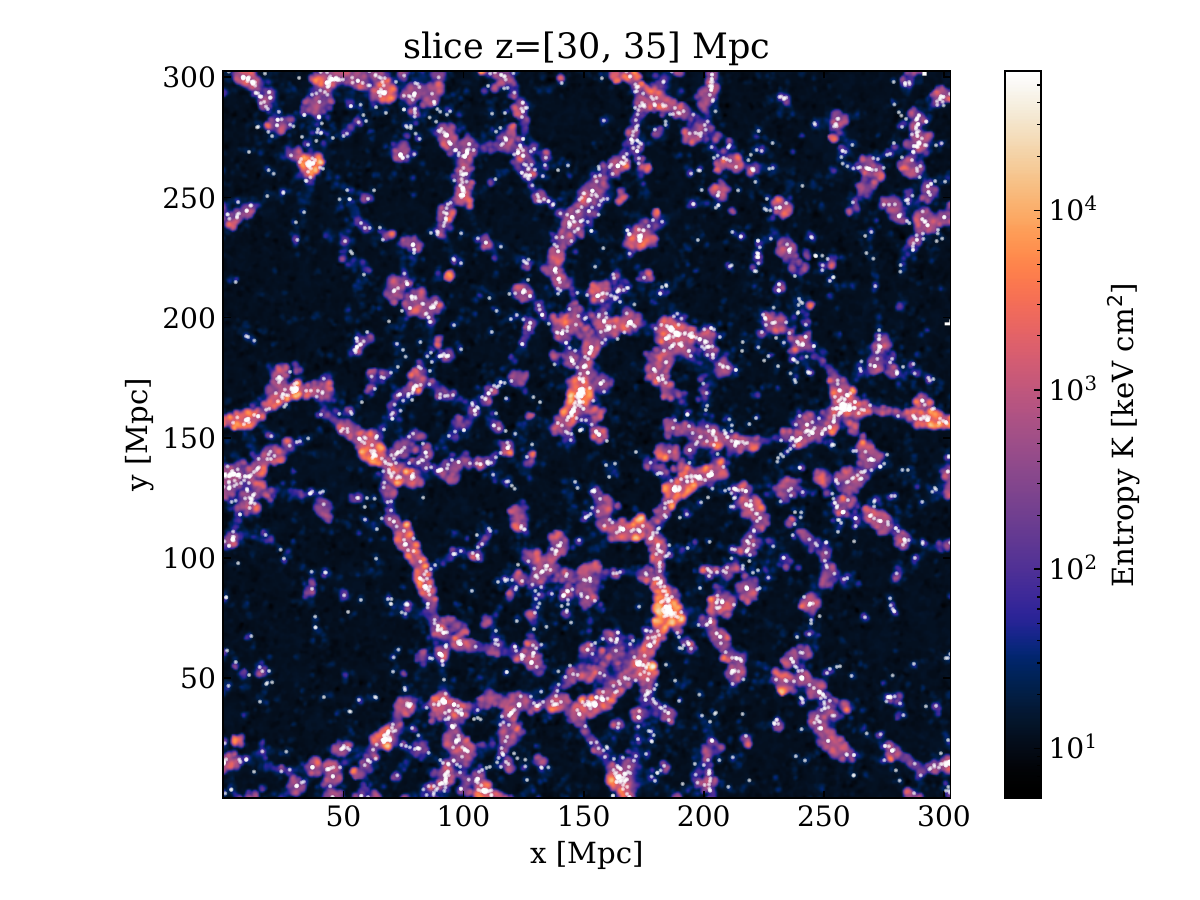}
    \includegraphics[width=1.\linewidth]{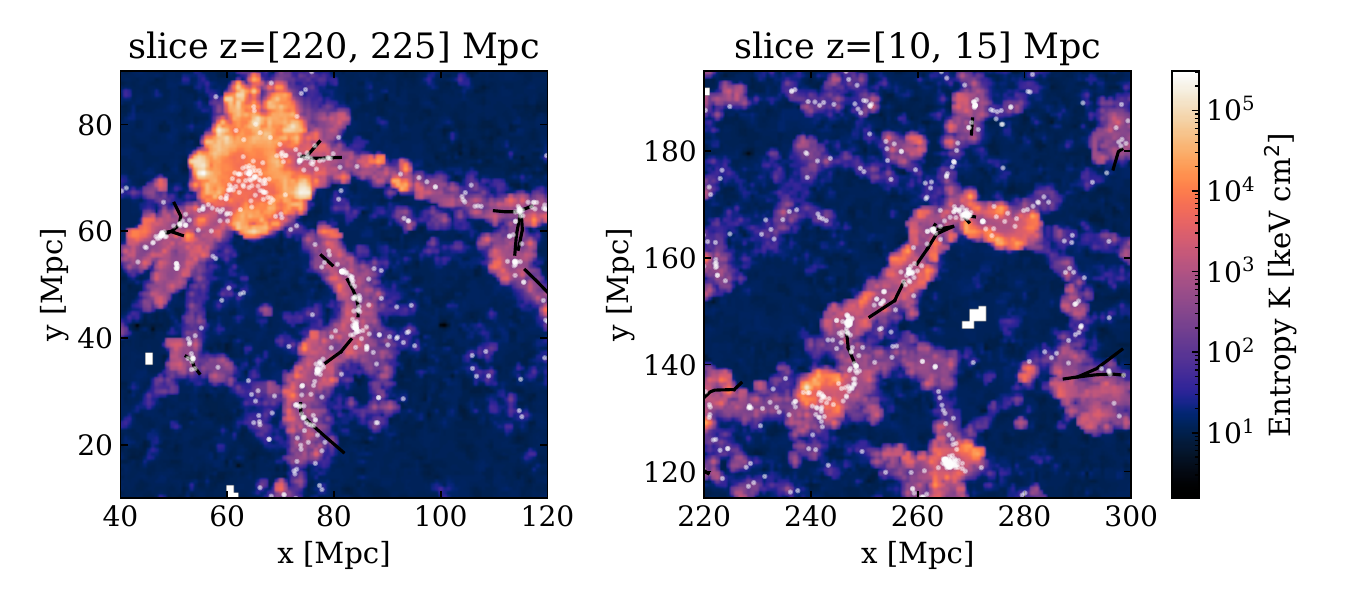}
    \caption{\textit{Top}: Gas entropy map in a thin slice of TNG300-1 at $z=0$. White dots mark $M_{*} > 10^{9} \, \mathrm{M}_\odot$ galaxies used to trace the filament network with DisPerSE. Entropy gradients delineate the cosmic web, with filamentary structures visible as elongated regions of higher entropy compared to the background. \textit{Bottom}: Zoom into two representative regions, with DisPerSE spines overlaid as black lines. The entropy increases from the inner to the outer parts of the filaments, producing a high-entropy boundary around the spine. This feature motivates the definition of $r_K$ as the radius of the entropy maximum (Sect.~\ref{SubSect:Rkcomputation}).}
    \label{Fig:Kmaps}
\end{figure*}

\subsection{Final sample}\label{SubSect:ResolutionCut}

Inspecting the gas density profiles of individual filaments, we find that some structures exhibit an offset of the density peak relative to the filament spine identified by DisPerSE. This is a caveat inherent to tracing filaments using galaxies. Because galaxies sparsely sample the underlying matter distribution, the reconstructed filament spine can be offset from the actual ridge of the underlying density field \citep{Bahe2025_filaments, Zakharova2023, Xu2026_filamentsDMpro}. Such misalignments are unavoidable in observational applications, where the underlying matter density field is not directly accessible. Therefore, to avoid introducing systematic biases in the filament property measurements, we discard these misaligned structures and retain only filaments whose density profiles peak at the DisPerSE spine. We have verified that this selection does not bias our final sample, since the discarded filaments neither exhibit distinctive gas properties nor preferentially occupy a particular region of the simulation volume. After removing the misaligned structures, $7587$ filaments remain.\\

To ensure that the radial profiles are not dominated by small-number statistics in the filament core, and that the entropy profile peak is robustly identified, we impose a conservative resolution requirement of at least $\sim10^3$ gas cells within the innermost radial bin. This threshold is based on gas-cell sampling tests as a function of radius and line-mass, which show a very rapid decrease in the number of cells towards the centres of the lowest line-mass filaments. We deliberately adopt the conservative value of $10^3$ cells to ensure that the innermost filament regions are sufficiently well sampled for robust statistics. This criterion corresponds to an effective line-mass of $\Lambda > 10^{10.8}\,\mathrm{M_\odot/Mpc}$, and filaments below this limit are excluded from the analysis because their central regions are insufficiently sampled from a statistical point of view.


Our final cut is designed to select high-contrast filaments, i.e., structures sufficiently prominent to be robustly traced in the galaxy distribution. Using the mean tracer density of the entire simulation box, $n_{\rm bkg}$, we compute the expected number of galaxies per unit length within a cylinder
of radius $R=1$ Mpc randomly placed in the simulation volume. This is given by $\lambda_{\rm bkg}=n_{\rm bkg} \, \pi \, R^2$.
For each filament, we compare this reference value to the number of tracer galaxies per unit length $\lambda$, whose computation is detailed in Sect.~\ref{Sect:Observables}.
Only filaments satisfying $\lambda >3 \,\lambda_{\rm bkg}$ are retained, thereby removing structures with a low galaxy contrast relative to the mean tracer density of the box. This leaves a final sample of $5637$ filaments.
We note that this cut is complementary to the $3\sigma$ persistence threshold adopted in the DisPerSE extraction. While the persistence threshold selects topologically significant features of the galaxy density field, it does not directly constrain the absolute number of galaxy tracers associated with an individual filament. A filament may therefore satisfy the persistence criterion while remaining sparsely sampled by galaxies. Our additional cut on $\lambda$ thus ensures that the selected structures are not only topologically significant but also sufficiently well traced by galaxies.\\

The successive selection steps reduce the initial catalogue of $10115$ filaments to $9019$ after 
node masking, to $7587$ after removing spine-offset structures, and to a final sample of $5637$ after applying the resolution and contrast thresholds described above. By construction, the analysis that follows will therefore focus on a conservative sample of prominent, high-contrast, and well-resolved filaments that can, in principle, also be robustly traced by galaxies using large-scale surveys.

\section{The entropy radius and gaseous line-mass}
\label{Sect:LambdaAndRadius}


Defining the boundary of a filament is not straightforward because different definitions and tracers can yield different characteristic radii. As for DM haloes, there is no unique preferred boundary, and the most appropriate radius often depends on the scientific question being addressed. Even in galaxy clusters, the inferred boundary has been shown to depend on the tracer and accretion geometry, particularly in anisotropic environments dominated by filamentary infall \citep[e.g.][]{Lebeau2024_sp}.

In this work we define the filament boundary using a thermodynamic criterion based on the radial entropy profile, rather than a fixed aperture or a density threshold. More specifically, we identify the boundary with the radius at which the entropy profile reaches its maximum, and refer to this scale as the entropy radius $r_K$. 


\subsection{Motivation: entropy as a filament boundary tracer}\label{Subsect:Motivation}

Figure~\ref{Fig:Kmaps} presents entropy maps for different thin slices of the simulation box. In the top panel, which shows the entire simulation box, we see that entropy gradients clearly delineate the presence of large-scale structure, particularly massive nodes and filaments. For reference, the $M_* > 10^{9}$ $\Msun~$ galaxies, used to trace the filaments with DisPerSE, are overplotted as white dots. The lower panels of Fig.~\ref{Fig:Kmaps} show zoomed-in regions, with the corresponding DisPerSE spines shown as black lines. Some spines are interrupted around massive haloes as a consequence of the masking procedure presented in Sect.~\ref{SubSect:filament_Cat}.  We notice a clear alignment between the filamentary structures and the entropy maps. The zoomed regions already reveal a radial entropy structure around the filaments, with lower values towards their cores and peaks near their boundaries.

Entropy traces the thermal history of the gas and is particularly sensitive to non-adiabatic processes such as shocks and mixing. As gas accretes onto structures, it can be decelerated and shock-heated, converting kinetic into thermal energy and producing sharp changes in temperature, pressure, and entropy (see Eq.~\ref{Eq:Entropy}). Entropy profiles can therefore identify transitions in the thermodynamic state of the gas. In galaxy clusters, minima in the entropy gradient have been associated with accretion shocks \citep{Aung2021_shocksplash,Towler2024,Lebeau2024_sp}, while maxima in the gradient occur closer to the splashback radius inferred from the gas density profile \citep{Towler2024,Lebeau2024_sp}. More recently, \citet{Sen2026_1halo2boundaries} used the entropy maximum itself as a proxy for the gas accretion-shock location. These studies show that different features of the entropy profile can trace distinct physical transitions.

Entropy also provides a possible connection to observations. Although it is not directly observed, it can be inferred from constraints on gas density and temperature, as demonstrated in galaxy clusters using X-ray data \citep[e.g.,][]{Pratt2010,Zhu2021_entropy_clusters}. Extending such measurements to cosmic filaments will be considerably more challenging because of their lower surface brightness, but may become feasible with future X-ray facilities such as NewAthena.

These considerations motivate our entropy-based definition of the filament boundary. We define the entropy radius, $r_K$, as the radial distance from the filament spine at which the entropy profile reaches its maximum. This provides a physically motivated transition between the filament gas and its surrounding environment, while being determined independently for each filament and therefore adapting to the diversity of their thermodynamic structures. We use the entropy maximum itself rather than extrema of its gradient, since derivatives are substantially more sensitive to noise in individual filament profiles. Finally, because entropy can in principle be reconstructed from observational constraints on gas density and temperature, this definition also offers a possible route towards measuring comparable thermodynamic boundaries in observed filaments.

\subsection{The entropy radius $r_K$}
\label{SubSect:Rkcomputation}

\begin{figure}
    \centering
    \includegraphics[width=1\linewidth]{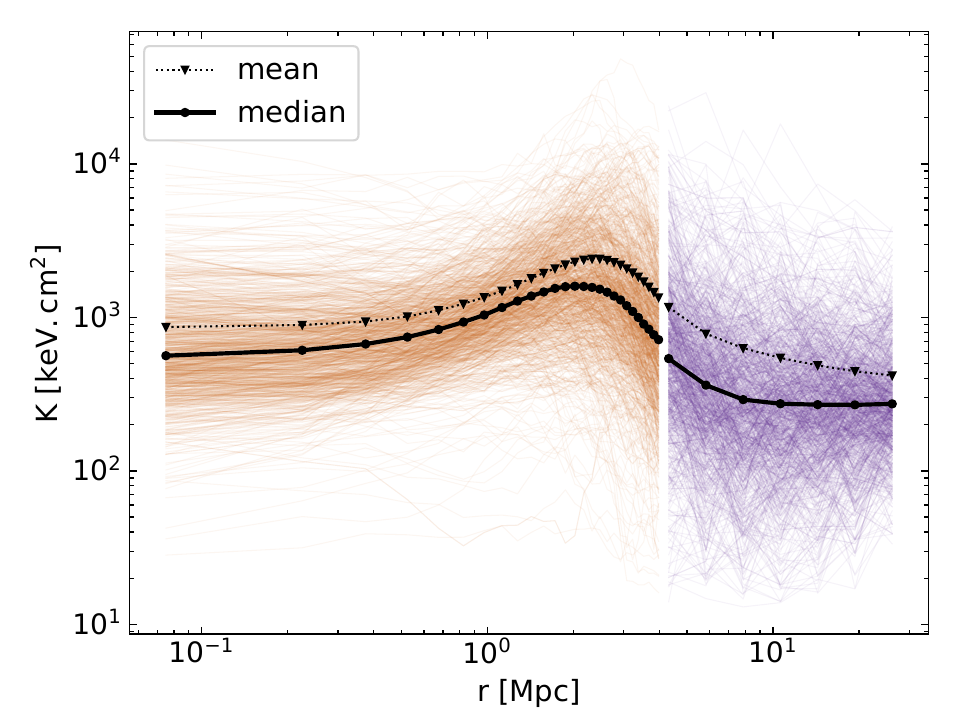}
    \caption{Entropy radial profiles of 1000 randomly selected filaments. The \textit{x}-axis corresponds to the cylindrical radius, perpendicular to the spine of the filament. The orange and purple curves respectively show results using the full gas cell dataset, and a randomly undersampled set of 1/500 cells. This differentiation was performed in order to capture the regions closer to filaments ($r< 4$~Mpc) in all details, while also providing the larger scale context and background. The profiles are not smoothed. We observe a clear bump in the entropy profiles. Note the logarithmic x-axis scale spanning a large 
    dynamic range. These individual filament profiles demonstrate that the entropy peak at $r_K$ (Fig.~\ref{Fig:THERMOpro}) is a consistent feature across the full population 
    and is not an artifact of stacking.}
    \label{Fig:Entropy_profiles1}
\end{figure}

\begin{figure}
    \centering
    \includegraphics[width=1\linewidth]{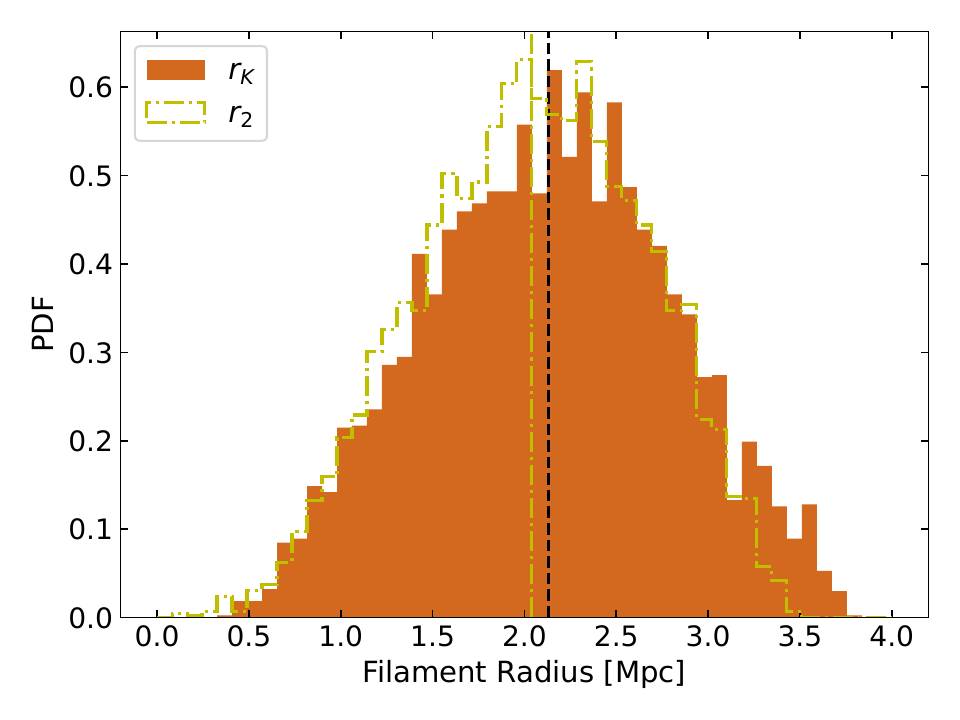}
    \caption{Filament entropy $r_K$ radius distribution (orange 
    filled histogram), derived in Sect.~\ref{SubSect:Rkcomputation}. 
    The black vertical line marks the position of the mean value, 
    $r_K = 2.13$~Mpc. The green dashed-dotted distribution corresponds to the $r_2$ radii, with mean value 2.04~Mpc, discussed in Sect.~\ref{SubSect:RadiiDiscussion} and shown 
    here for comparison. The tails of the $r_K$ distribution reflect the physical diversity of the filament population: 
    filaments with $r_K \lesssim 0.5$~Mpc are predominantly low-$\Lambda$ structures in underdense environments, while 
    those with $r_K \gtrsim 3.5$~Mpc are high-$\Lambda$ filaments in denser large-scale environments near massive nodes (this is explored quantitatively in 
    Sect.~\ref{SubSect:StructuralCorr}).}
    \label{Fig:Radii_dist}
\end{figure}

In Fig.~\ref{Fig:Entropy_profiles1} we present the radial entropy profiles of individual filaments. At small radial distances ($r<4$~Mpc), shown in orange, the profiles are computed using the full gas-cell dataset and the method introduced in Sect.~\ref{SubSect:Method_Profiles}. At larger radii, shown in purple, we use a randomly undersampled gas catalogue containing 1/500 of the cells to reduce the computational cost. These outer regions are included only for visualisation purposes. The profiles exhibit a clear enhancement in entropy at distances of $r \sim 1$ -- $3$~Mpc from the filament spine. This feature is also visible in the median profile, shown by the thick solid line with dots. At its peak, the median entropy increases by $488\%$ relative to the background entropy level, defined as the mean value of the median entropy profile at large distances ($r>10$~Mpc).
This factor of $\sim 6$ entropy enhancement identifies a characteristic scale in the thermodynamic structure of filament gas. Motivated by studies of gas entropy around galaxy clusters \citep[e.g.,][]{Voit2005_entropy, Lebeau2024_sp, Sen2026_1halo2boundaries}, we consider accretion shocks as one possible physical origin of this feature. Indeed, visual inspection of individual filaments reveals sharp features in their temperature and electron-density profiles in some cases, potentially consistent with shocks.
Alternatively, the entropy maximum may reflect radial changes in the relative contributions of dense and diffuse gas. Previous work using
TNG300-1 showed that the WHIM accounts for approximately $80\%$ of
the gas mass around $r\sim 1-2$ Mpc, while denser and hotter phases become
increasingly important, although remaining subdominant, towards
the filament spine \citep{GalarragaEspinosa2021}. Therefore, since $K \propto T\,n_e^{-2/3}$ is computed cell by cell before volume averaging within a bin, a greater contribution from diffuse, higher-entropy gas could enhance the mean entropy at intermediate radii, while an increasing contribution from dense, lower-entropy gas towards the spine could explain the decline towards the filament core (see also Sect.~\ref{Subsect:TmTv_ratio}).
Accretion shocks and variations in gas-phase composition may both contribute to the observed entropy peak. Distinguishing their relative importance would require a dedicated analysis of multi-phase gas dynamics and structure, which we leave for future work. Here, we adopt the entropy peak as a physically motivated definition of the filament radius.

The filament entropy radius, $r_K$ is therefore defined as the radius at which the gas entropy reaches its maximum value. The entropy peak is identified for each filament individually from its smoothed entropy profile to reduce the impact of small-scale noise. For each filament, we first quadratically interpolate its $\log(K)$ profile to increase the number of sampling points. We then apply a Savitzky--Golay filter\footnote{From the Python \texttt{scipy.signal} library \citep{Virtanen2020SciPy}.} with a window of 7 bins ($7 \times 150$~kpc) and a second-order polynomial. These parameters were optimised to preserve the entropy peak while smoothing small-scale noise; the procedure is robust across the full range of profile shapes present in the population, from narrow well-defined peaks to broader transitions, as reflected in the diversity of individual profiles visible in Fig.~\ref{Fig:Entropy_profiles1}.
The resulting distribution of $r_K$ is shown in Fig.~\ref{Fig:Radii_dist}. It is approximately symmetric and centred around $\sim 2$~Mpc, with a mean value of $r_K = 2.13$~Mpc and a range of $0.3$--$3.7$~Mpc.

\subsection{Comparison with other radius definitions}
\label{SubSect:RadiiDiscussion}

The width of cosmic filaments has been defined in several ways in the literature, using different matter tracers and physical criteria. As discussed above, this reflects the absence of a unique filament boundary and the fact that different definitions probe different aspects of these structures.

Inspired by halo studies, one possible approach is to define a filament radius using a fixed density threshold. Unlike haloes, however, for which overdensity radii such as $R_{200c}$ are commonly adopted, filaments are not fully virialised structures, and there is no obvious universal density threshold defining their boundary. Nevertheless, for comparison, we compute a density-based radius $r_2$, defined as the radius at which the electron density decreases to twice its local floor level. For each filament, the floor level is measured using their $n_e$ profile at distances of $3.5$--$4$~Mpc from the spine. The resulting $r_2$ distribution, shown by the green dash-dotted curve in Fig.~\ref{Fig:Radii_dist}, spans a similar range to the $r_K$ distribution. At the individual filament level, however, we find that the two radii are not correlated.

Other density-based definitions have been proposed. For example, \citet{YangQirui2025_width_filaments} defines filament widths from the radius at which the gas density approaches the mean cosmic gas density, obtaining characteristic scales broadly comparable to those measured here. However, such a definition does not account for variations in the large-scale environment surrounding individual filaments. To address this, \citet{Bahe2025_filaments} instead defines the filament boundary of the DM density relative to a locally measured background, identifying the radius at which the density falls below twice 
this local floor. Although the factor of two is arbitrary, this definition captures the point at which the filament merges into its immediate environment. They find characteristic DM filament radii of the order of 1~Mpc, with values spanning approximately $0.5$--$3$~Mpc at $z=0$. While this range is consistent with our findings, the mean radius is lower than our gas-based estimate, as expected from the distinct dynamics of collisionless dark matter and gas.

A complementary approach is to define filament boundaries dynamically. In the collisionless DM component, filament formation is associated with shell crossing and the folding of the dark-matter phase-space sheet, producing multistream regions bounded by caustics \citep{Neyrinck2012_Origami,Feldbrugge2018,Feldbrugge2025}. Motivated by the analogy with the splashback radius of DM haloes, \citet{Wang2024fil_radius} defined a characteristic 
filament radius from the minimum logarithmic slope of the radial galaxy number-density profile. Using stacked profiles, they find a characteristic radius of approximately $1$~Mpc at $z=0$ in both simulations and observations. This approach is well suited to determining an average filament boundary, but is difficult to apply to individual structures because the discrete galaxy distribution produces noisy individual density profiles and derivatives. Measuring splashback radii for individual filaments would require a dedicated analysis of the position-velocity phase space of DM particles, which is beyond the scope of this work.

Another approach is to characterise filament widths by fitting analytic models to their average radial density profiles. For example, \citet{GalarragaEspinosa2020} fitted Einasto, $\beta$, and NFW models to the average galaxy number density profiles of filaments and obtained (model dependent) characteristic radial scales of approximately $0.14$--$0.25$ Mpc. These scales are useful for describing the shape of the density profile, but their physical interpretation as filament boundaries is less direct, since the fitted functional forms are empirical and are not derived from a physical model of filament formation. They should therefore be interpreted primarily as characteristic scales marking changes in the profile shape, rather than as physically defined filament edges.\\

In summary, filament radii depend on both the tracer and the physical definition adopted. Different estimators can yield similar characteristic scales at the population level, while identifying different boundaries for individual filaments. Among the definitions considered, we adopt the entropy radius $r_K$ throughout this work. It provides a physically motivated boundary defined individually from the thermodynamic structure of each filament. Although entropy is not itself a direct observable, it can be inferred from reconstructed gas density and temperature profiles, as routinely done for galaxy clusters \citep[e.g.,][]{Pratt2010,Zhu2021_entropy_clusters}, using X-ray and SZ observations. The other filament radius definitions discussed above are presented for comparison but are not used in the line-mass computation.

Finally, it is important to recall that filaments are not perfect cylinders with a constant radius along their spine. Their width can vary along their length 
\citep{YangQirui2025_width_filaments} and with distance to connected nodes and clusters \citep{Wang2024fil_radius}. The radius assigned to each filament in this work should therefore be interpreted as the mean characteristic scale along its length.

\subsection{Definition and measurement of $\Lambda$}
\label{SubSect:LambdaDef}

Having defined a characteristic radius for each filament, we can now quantify its gas mass content. 
A natural quantity to characterise the mass content of an elongated structure is its linear mass density, $\Lambda$, i.e. the mass per unit length, also called `line-mass' for brevity. We define it as:
\begin{equation}\label{Eq:Lambda_def}
    \Lambda = \frac{M(r < r_\mathrm{fil})}{L_\mathrm{fil}},
\end{equation}
where $M(r<r_\mathrm{fil})$ is the mass enclosed within a filament of radius $r_\mathrm{fil}$ and length $L_\mathrm{fil}$.
In this work, we focus on the gas component and adopt the entropy radius introduced in Sect.~\ref{Sect:LambdaAndRadius}, such that $r_\mathrm{fil}=r_K$. Considering cylindrical geometry and that the gas density varies only radially, with $\rho_{\rm gas}(r) = \mu_e\,m_p\,n_e(r)$, the enclosed gas linear mass density reduces to the radial integral
\begin{equation}\label{Eq:Lamba_ne}
    \Lambda = 2\pi\,\mu_e\,m_p 
    \int_0^{r_K} r \, n_e(r) \, dr.
\end{equation}
We note that the electron density profile $n_e(r)$ includes all gas cells retained after the masking procedure (described in Sect.~\ref{SubSect:Method_Profiles}), without any additional selection by temperature or gas phase. Thus, $\Lambda$ measures the total gas mass within $r_K$, not only the diffuse or WHIM component.

\subsection{The filament line-mass function}
\label{SubSect:LineMassFunction}

\begin{figure}
    \centering
    \includegraphics[width=1\linewidth]
    {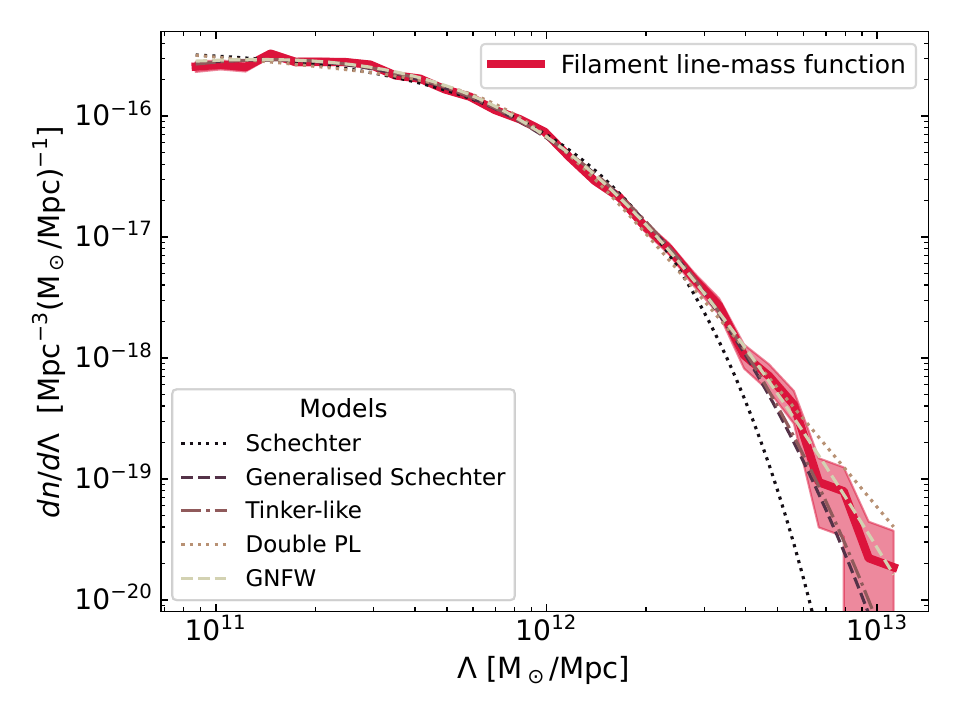}
    \caption{Gaseous line-mass function of cosmic 
    filaments measured in TNG300-1 (red curve). 
    Error bands correspond to Poisson uncertainties on the filament counts in each line-mass bin, quantifying the statistical uncertainty associated with the finite simulation volume. 
    The other curves show the best-fit curves of empirical models taken from halo literature
    (see Sect.~\ref{SubSect:LineMassFunction} and Appendix~\ref{APP:MCMC}).}
    \label{Fig:Lambda_fils}
\end{figure}

We compute $\Lambda$ for each filament by applying Eq.~\ref{Eq:Lamba_ne} using the individual electron density profiles and entropy radii $r_K$ derived in Sect.~\ref{SubSect:Rkcomputation}. The resulting line-mass function is shown by the solid red curve in Fig.~\ref{Fig:Lambda_fils}. To our knowledge, this is the first measurement of the gaseous line-mass function of cosmic filaments from a cosmological hydrodynamical simulation. This distribution, consisting of the number density of filaments per unit interval in $\Lambda$, provides a statistical baseline for future observational surveys and a benchmark for semi-analytic and analytical models of large-scale filament formation. Its shape encodes information about the distribution of large-scale overdensities from which filaments form, and its high-$\Lambda$ end is a sensitive probe of the rarest and most massive structures in the cosmic web.

The distribution shows a relatively flat regime at $\Lambda \lesssim 5 \times 10^{11}\,\Msun{}\,\mathrm{Mpc}^{-1}$, where the bulk of the population resides, followed by a steep decline towards higher line-masses. This flat low-$\Lambda$ regime may partly reflect the anisotropic growth of filaments, which accrete matter predominantly towards their spines while redistributing it along their length, allowing structures with different total gas masses and lengths to reach similar $\Lambda$ values.\\

To characterise the shape of the line-mass function, we fit a sequence of increasingly flexible empirical functional forms using Markov chain Monte Carlo (MCMC) with a Gaussian likelihood. A simple Schechter function ---combining a power law with an exponential cutoff--- reproduces the bulk of the distribution but underestimates the abundance of the highest $\Lambda$ filaments ($\chi^2_\nu = 3.93$). A generalised Schechter function with a free suppression index $\beta$ improves the fit considerably ($\chi^2_\nu = 1.17$). The best description ($\chi^2_\nu = 1.09$) is provided by a Tinker-like functional form 
\citep{Tinker2008}:
\begin{equation}\label{Eq:Tinker}
    \frac{dn}{d\Lambda} = C 
    \left[1 + \left(\frac{\Lambda}{\Lambda_c}
    \right)^\alpha\right]
    \exp\!\left[-\left(\frac{\Lambda}{\Lambda_c}
    \right)^\beta\right],
\end{equation}
where $C$, $\Lambda_c$, $\alpha$, and $\beta$ are free parameters. We also tested a double power-law model ($\chi^2_\nu = 2.74$) and a generalized NFW profile \citep[GNFW;][]{Navarro1997_NFW}, with $\chi^2_\nu = 1.13$. Neither improves on the Tinker-like description, with the former failing to reproduce the rapid high line-mass decline and the latter introducing additional parameter degeneracies without a meaningful gain in fit quality.
All the details on the fit results can be found in Appendix~\ref{APP:MCMC}.

The success of the Tinker-like functional form is physically suggestive. The exponential suppression of the line-mass function at high $\Lambda$ mirrors the exponential suppression of the halo mass function at high masses, both reflecting the rarity of extreme overdensities in a Gaussian random field. Whether this analogy extends to a deeper structural connection between the formation of massive haloes and that of high-$\Lambda$ filaments is a question we leave for future work. \\

We note that these functional forms are purely phenomenological and are not derived from a physical model of filament formation or evolution. A deeper theoretical study will be needed to connect the line-mass function to the underlying processes governing filament growth from first principles. 
Also, we note that cosmic variance might affect the abundance of the high line-mass filaments, given the finite simulation volume of the TNG300 box. Larger cosmological volumes will be useful to further quantify the high-$\Lambda$ tail.

Furthermore, while we focus here on the gas component, a complete description should ultimately include the dark matter contribution. Previous work has shown that gas broadly follows the dark matter distribution in filaments but with systematic radial offsets, including a gas depletion towards the filament core and an excess at ${\sim}\,1.5\,\mathrm{Mpc}$ from the spine \citep{GalarragaEspinosa2022}. Extending the present analysis to the total matter content will be important for connecting the gas line-mass function to the underlying dark matter distribution.

The line-mass function derived here provides a direct statistical prediction for observational surveys: combined with the richness proxy calibration (Sect.~\ref{Sect:Observables}), it provides a measurable observable for current SDSS and forthcoming PFS and DESI data. Having measured $\Lambda$ for each filament in our catalogue, we now investigate how it connects to the structural and thermodynamic filament properties in the next section.

\section{Filament structure and thermodynamics}
\label{Sect:FilProperties}

Having characterised the entropy radius and line-mass of individual filaments in Sect.~\ref{Sect:LambdaAndRadius}, we now investigate how $\Lambda$ connects to filament morphology and gas thermodynamic state.

\subsection{Line-mass--morphology correlations: $\Lambda$, $r_K$, $L$}
\label{SubSect:StructuralCorr}

\begin{figure}
    \centering
    \includegraphics[width=1\linewidth]{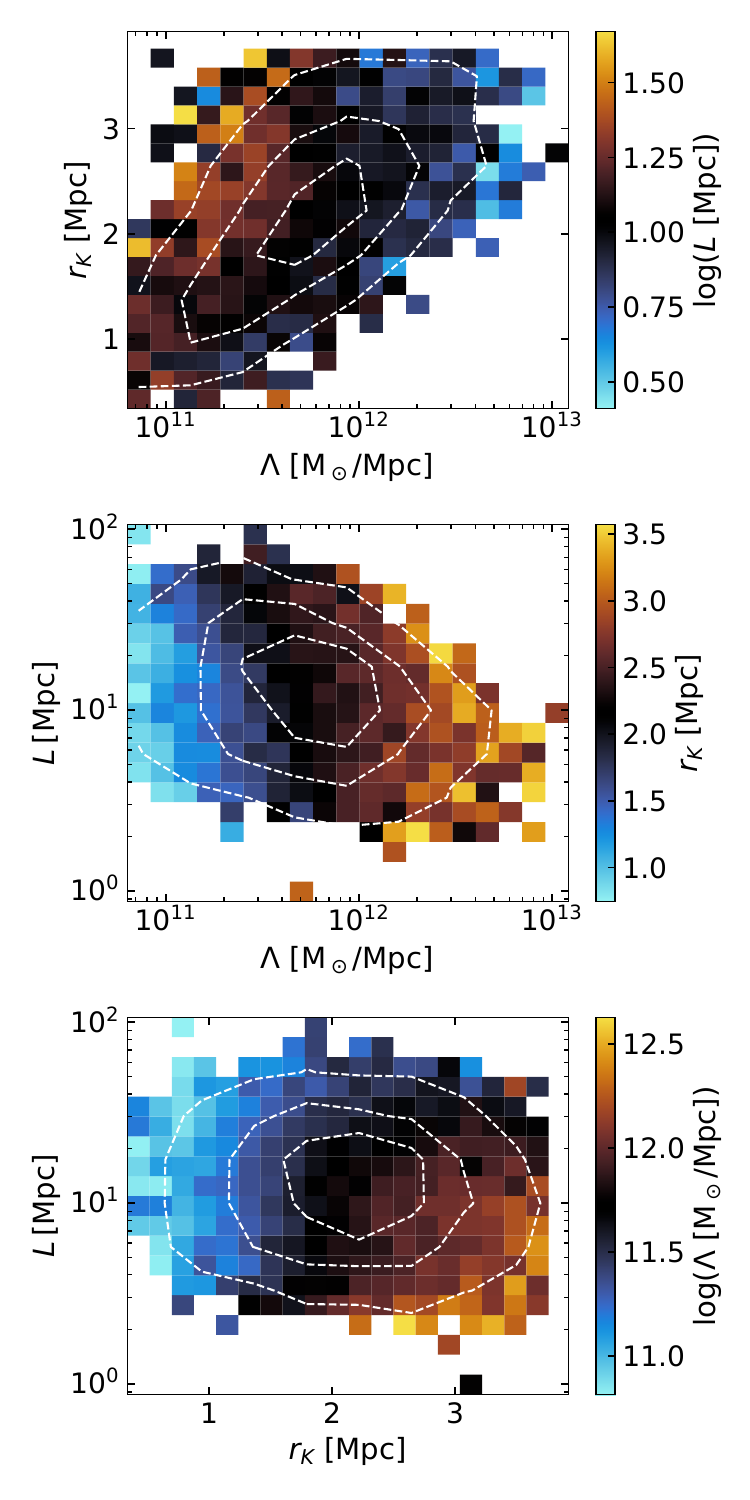}
    \caption{Projections of the 3D parameter space of filament properties, namely line-mass vs radius, line-mass vs length, and radius vs length, from top to bottom panels. Pixel colours correspond to the median value of the other parameter within the corresponding $x-y$ bin. From inner to outer, the contours show the $33\%$, $68\%$ and $95\%$ percentiles of the data.}
    \label{Fig:binnedstatistics}
\end{figure}

We investigate the intrinsic correlations between filament line-mass, radius and length, and demonstrate that $\Lambda$ organises the diversity of the filament population more effectively than filament length alone. Filament length, $L$, is computed by summing the lengths of the individual segments composing each filament, following the filament path identified by DisPerSE \citep[see e.g.,][for details]{GalarragaEspinosa2020, GalarragaEspinosa2024_MTNG}.\\

Figure~\ref{Fig:binnedstatistics} shows the three projections of the $\Lambda$-$r_K$-$L$ parameter space. The colour scale traces the third parameter, allowing us to visualise the joint dependence of these quantities. The contours enclose the $95\%$, $68\%$, and $33\%$ percentiles.
The top panel of this figure shows a strong correlation between line-mass $\Lambda$ and radius $r_K$, with a Pearson coefficient of $0.68$. 
This behaviour is consistent with the electron density profiles (not shown for conciseness), where thicker filaments have systematically higher $n_e$ values across the full radial range, resulting in larger enclosed gas masses and, therefore, higher $\Lambda$. 

The correlation between filament line-mass and length is milder but also significant (middle panel). A Pearson correlation coefficient of $-0.30$ indicates that filaments with higher $\Lambda$ tend to be shorter. This is consistent with previous findings that short filaments preferentially reside in denser large-scale environments, closer to massive nodes, \citep{GalarragaEspinosa2020, GalarragaEspinosa2022} where accretion rates are higher \citep{Borzyszkowski2017_ZOMG1}, potentially increasing the gas mass accumulated per unit length. 
We note, however, that this relation shows a large scatter, complicating a simple modelling of the line-mass--length relation.
We verified that this trend is not simply driven by a geometrical scaling or volume effect by comparing the effective mean gas density within $r_K$, $\bar{\rho}\equiv\Lambda/(\pi r_K^2)$, with the expectation $\bar{\rho}\propto L^{-1}$. We find that $\bar{\rho}$ shows a much weaker dependence on $L$ than this expected slope of $-1$. We also note that the estimator used for $\Lambda$ in Eq.~\ref{Eq:Lamba_ne} is obtained directly from the radial electron-density profile and therefore contains no explicit dependence on $L$.
From an evolutionary perspective, however, an increase in $\Lambda$ for an individual filament can still result from a decrease in its length (e.g. through longitudinal contraction), an increase in gas mass (e.g. through radial accretion), or a combination of both. Distinguishing between these different assembly pathways would require tracking individual filaments and their gas flows across cosmic time, which is beyond the scope of this work.

We find a mild negative correlation between filament length and radius (Pearson coefficient of $-0.10$, bottom panel). The weakness of this correlation, together with the large scatter, indicates that filament length and radius capture independent aspects of filament structure: filaments of similar length can in general span a wide range of radii, and vice versa.
Line-mass, radius and length describe together the diversity of the filament population.
We find that the median filament in our sample has a line-mass $\Lambda = 10^{11.67} \Msun{}/\mathrm{Mpc}$, a radius $r_K = 2.15$~Mpc (mean $2.13$~Mpc; Fig.~\ref{Fig:Radii_dist}), and a length $L = 11.44$~Mpc.

Now focusing on the colour scale, these three projections show a clear continuous sequence from predominantly long, thin, low-$\Lambda$ filaments to shorter, wider, high-$\Lambda$ structures. In particular, the bottom panel shows that filament morphology varies systematically with $\Lambda$, demonstrating that line-mass is an organising parameter of filament morphology. 
In what follows, we use $\Lambda$ as the primary property to investigate whether this diversity also extends to the physical properties of the filament gas.

\subsection{Thermodynamic profiles ordered by $\Lambda$}
\label{SubSect:ThermoPro}

\begin{figure*}
    \centering
    \includegraphics[width=1\linewidth]{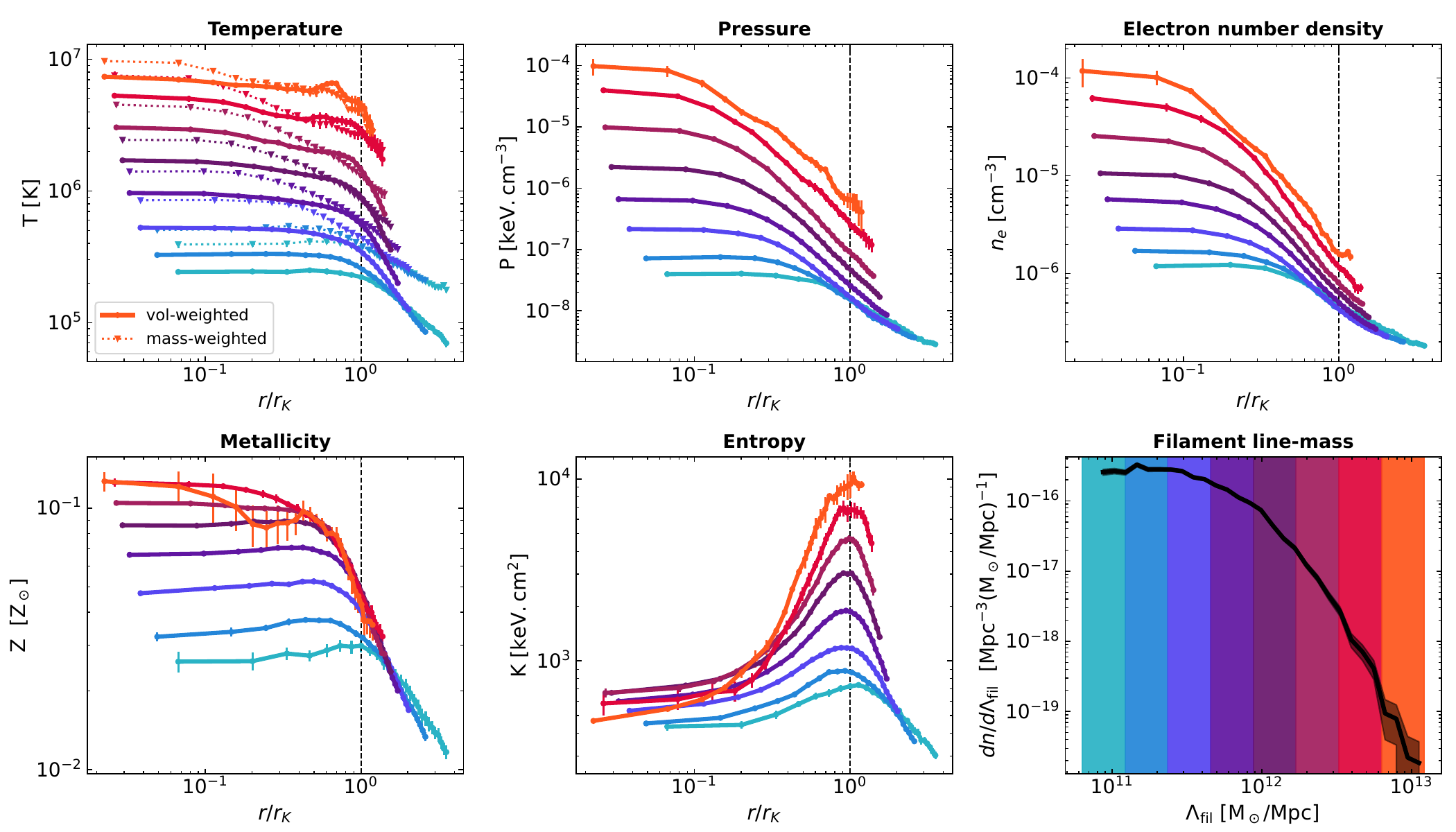}
    \caption{Radial filament profiles of different gas properties, as a function of filament line-mass $\Lambda$. Curves correspond to the median values within each bin, and errors on the median are estimated using the bootstrap method. The radial coordinates are rescaled by the median $r_K$ value within each bin. From lowest to highest $\Lambda$, the number of filaments in each bin is 451, 916, 1399, 1483, 962, 345, 74, and 7. The highest-$\Lambda$ bin (7 filaments) is shown for completeness but should be interpreted with caution given the limited statistics.    }
    \label{Fig:THERMOpro}
\end{figure*}

Having established that filament morphology varies systematically with $\Lambda$, we now ask whether the physical state of the gas follows the same ordering. Figure~\ref{Fig:THERMOpro} presents the radial profiles of the different gas properties, separated into thin bins of filament line-mass. Bins are linearly separated in log-space. Radial distances are normalised by the median $r_K$ value within each bin, allowing us to compare the internal structure of filaments of different widths. A clear ordering with $\Lambda$ emerges across all the quantities considered, with higher-$\Lambda$ filaments hosting hotter, denser, higher-pressure, and more metal-enriched gas compared to lower line-mass filaments.

The temperature profiles show one of the strongest dependencies on line-mass. Low line-mass filaments have core temperatures of a few $10^5$ K, while structures with the highest line-masses reach almost $10^7$ K, confirming that the filament population is squarely placed within the WHIM temperature window \citep[$10^5$--$10^7$ K;][]{Dave2001_massing_baryons, CenOstriker2006}. Within $r_K$, the profiles are relatively flat and clearly separate according to $\Lambda$, before decreasing outside the filament boundary. 
This behaviour is consistent with our previous TNG300 results, where the gas temperature was found to depend on filament length \citep{GalarragaEspinosa2021}. Here, however, separating filaments by line-mass reveals a substantially larger variation in temperature, suggesting that $\Lambda$ captures the thermodynamic diversity of the filament population more effectively than length. Similar trends have been reported in other hydrodynamical simulations, in which denser and more prominent filaments contain systematically hotter gas \citep[e.g.,][]{GhellerVazza2019_surveyTandNTprops_fils,Tuominen2021}.

The pressure and electron-density profiles show a similarly strong dependence on the line-mass, with both quantities increasing systematically with $\Lambda$ across the filament population. The pressure variation is particularly pronounced, reflecting the combined increase in gas density and temperature towards higher line-masses. These trends further show that $\Lambda$ captures large differences in the thermodynamic state of filament gas.

A clear dependence on line-mass is also found for metallicity. The cores of filaments with the highest line-masses reach $Z\sim0.1$--$0.13\,Z_\odot$, compared to $Z\sim0.03\,Z_\odot$ for the lowest line-mass structures, indicating substantially stronger chemical enrichment in those structures. This likely reflects the cumulative interaction between the galaxies residing in these environments and the surrounding filament gas. We note, however, that metal enrichment in cosmological simulations depends on the adopted stellar evolution and feedback prescriptions, and the absolute values reported here should therefore be interpreted within the context of the TNG model. Nevertheless, our metallicity estimates for cosmic filaments are consistent with the low values generally expected for the WHIM \citep[e.g.,][in the EAGLE simulations]{Tuominen2021} and with recent X-ray measurements of individual massive filaments, which find metallicities of order $\lesssim0.1\,Z_\odot$ \citep[e.g.,][]{Veronica2024,Dietl2024}. 
Importantly, the enrichment measured here is not driven by gas associated with massive haloes or clusters, which is removed by the masking procedure described in Sect.~\ref{SubSect:Method_Profiles}. The trend is therefore present in the remaining filament gas and is consistent with large-scale metal transport driven by stellar and AGN feedback over cosmic time \citep[e.g.,][]{Springel2018_TNG,Pillepich2018_TNGgalaxies}. The flattening of metallicity at high $\Lambda$, around $Z\sim0.1$--$0.13\,Z_\odot$, may reflect a balance between metal enrichment and dilution by the accretion of lower-metallicity gas from the surrounding large-scale structure, although disentangling these effects is beyond the scope of this work.

Finally, the entropy profiles reveal a strong dependence of the thermodynamic boundary on the filament line-mass. While all profiles peak at $r/r_K\simeq1$ by construction, 
the entropy enhancement becomes increasingly
pronounced with $\Lambda$. 
High line-mass filaments exhibit a much stonger entropy values at their peak compared to low line-mass filaments. This may reflect more efficient shock heating of accreting gas in deeper filament potential wells, but also differences in gas-phase composition across the filament population, as will be briefly discussed next.

In summary, these profiles show that $\Lambda$ systematically organises the physical state of the filament gas. Beyond the morphological trends discussed in Sect.~\ref{SubSect:StructuralCorr}, filaments form a continuous thermodynamic sequence with line-mass, from relatively cool, diffuse, and metal-poor structures to hotter, denser, and more enriched filaments with a stronger entropy boundary. An interesting aspect to note is that filaments at the high-$\Lambda$ end have properties that are close to the thermodynamic conditions found in the outskirts of galaxy clusters \citep{Pratt2010}.

\subsection{Thermal inhomogeneity: $T_m$/$T_v$}
\label{Subsect:TmTv_ratio}

\begin{figure}
    \centering
    \includegraphics[width=1\linewidth]{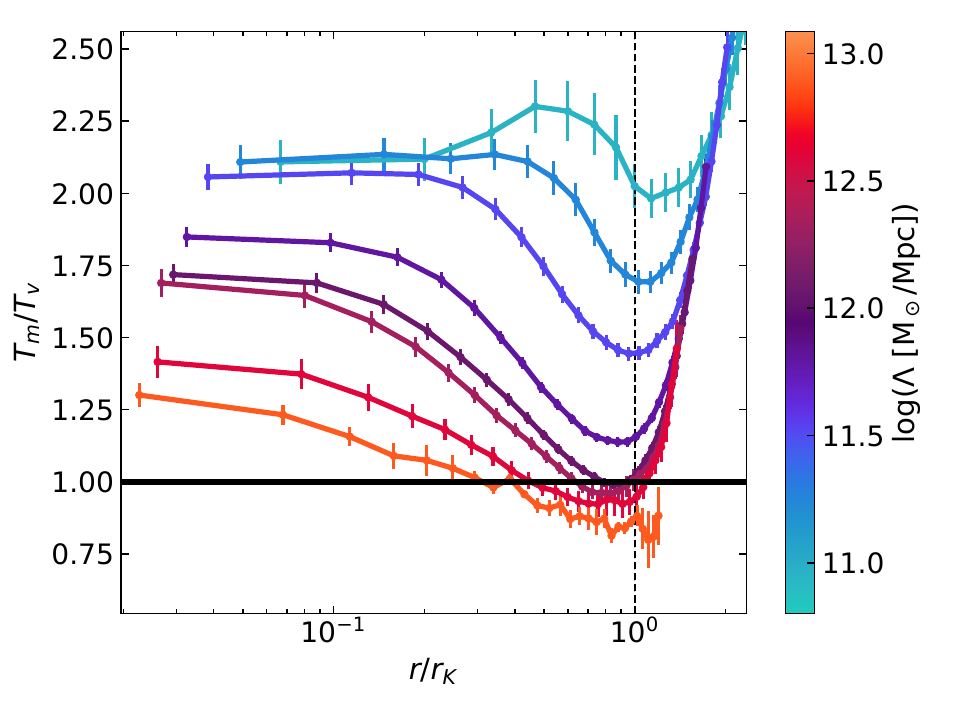}
    \caption{Ratio between the mass- and volume-weighted temperature profiles around filaments (defined in Sect.~\ref{SubSect:Method_Profiles}). This ratio quantifies the distribution of temperature between dense and diffuse gas components in filaments.}
    \label{Fig:TmTv}
\end{figure}

The previous results show that the average thermal state of the filament gas strongly depends on line-mass $\Lambda$. We now investigate how this thermal energy is distributed between gas components of different densities. To this end, we study the ratio between the mass- and volume-weighted temperatures, $T_m/T_v$, as a function of filament radius and line-mass. This ratio provides a simple diagnostic of thermal inhomogeneity of the gas: values above unity indicate that the denser gas is preferentially hotter, while $T_m/T_v<1$ indicates that the hotter gas is preferentially in a diffuse state. The resulting radial profiles are presented in Fig.~\ref{Fig:TmTv}.

The ratio of mass-weighted to volume-weighted temperature $T_m/T_v$ reveals a fundamental transition with $\Lambda$ in the internal thermal structure of filaments. In the lowest line-mass filaments (cyan), $T_m/T_v \sim 2$ throughout the inner regions, indicating that hot gas in these structures is preferentially associated with dense substructures ---such as galaxies and their surrounding haloes--- embedded in a cooler, more diffuse medium. Radiative cooling may contribute to this thermal inhomogeneity, and could potentially promote thermal instability, since the characteristic gas temperatures of these low-$\Lambda$ filaments lie near the regime where radiative cooling is particularly efficient \citep[e.g.][]{Mandelker2021}. However, $T_m/T_v$ should not be interpreted as a direct diagnostic of thermal instability, and the relevant cooling scales may be only marginally resolved in TNG300. We therefore leave a detailed analysis of cooling and dynamical timescales to future higher-resolution work.

As $\Lambda$ increases, $T_m/T_v$ decreases and falls below unity near $r_K$ in the highest line-mass filaments (orange), indicating that the diffuse gas becomes comparatively hotter. This behaviour is consistent with an increasing contribution from shock-heated accreting gas, although changes in gas-phase composition may also play a role, as discussed in Sect.~\ref{SubSect:Rkcomputation}.\\

This transition in filament structure with increasing line-mass, from a hot-dense to a hot-diffuse medium, has direct implications for the detectability of filament gas in X-ray and Sunyaev-Zel'dovich observations \citep[e.g.][]{DeGraaff2019, Tanimura2019, Tanimura2020_Rosat, Tanimura2022_eRosita, Veronica2024, Zhang2024erosita_fils}. X-rays preferentially trace the denser component through their $\sim n_e^2$ dependence, while the tSZ signal traces the integrated thermal pressure of both dense and diffuse gas. The relative sensitivity of X-ray and tSZ observations to filament gas is therefore expected to change with filament line-mass, as a reflection of how the thermal energy is distributed between dense and diffuse gas.

\subsection{Gas scaling relations with filament line-mass}\label{Subsect:gas_scalings}

\begin{figure*}
    \centering
    \includegraphics[width=1\linewidth]{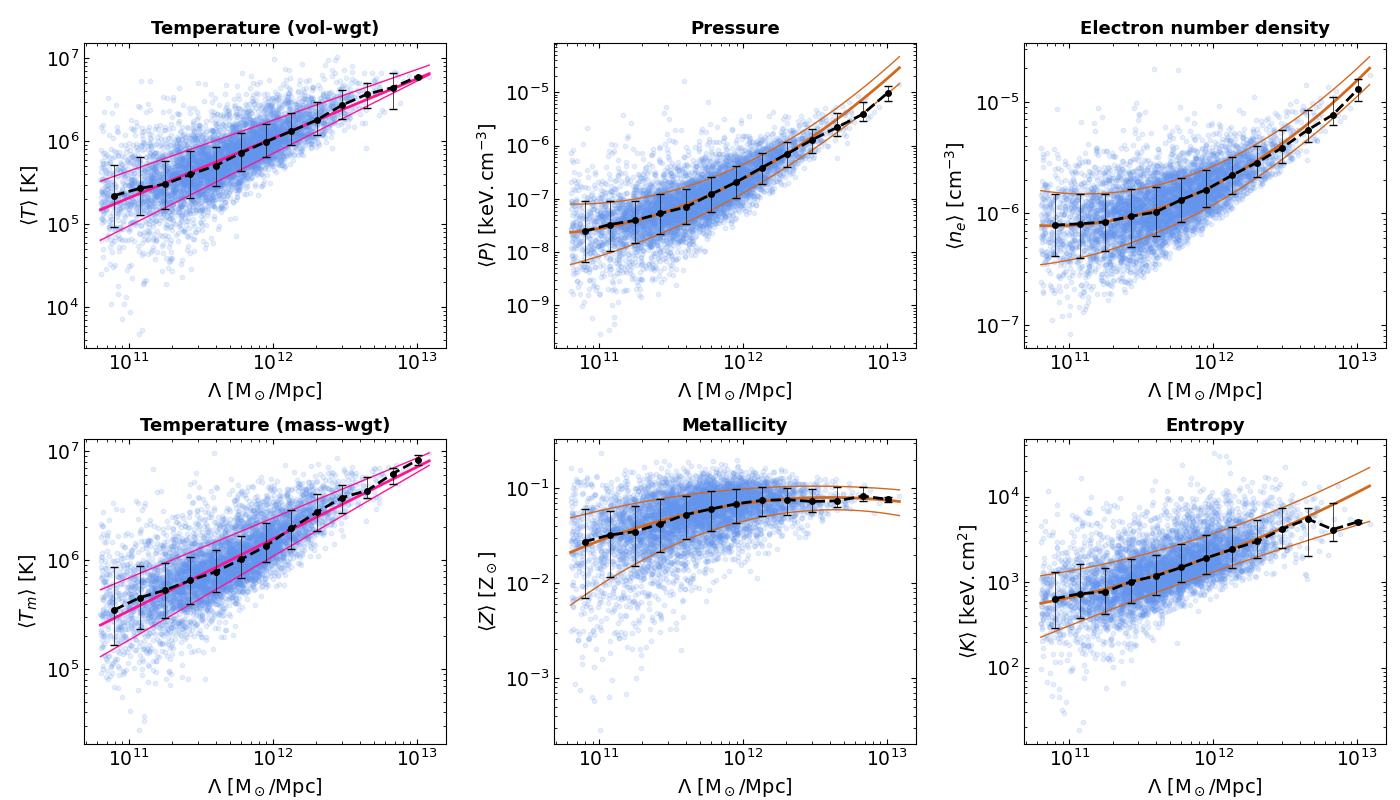}
\caption{Scaling relations between filament line-mass and gas properties. Each blue point represents an individual filament. Black dashed curves show the measured median relations, with error bars indicating the 16th and 84th percentiles. The pink and orange curves show the quantile-regression fits for the linear (Eq.~\ref{Eq:lin_regression}) and quadratic (Eq.~\ref{Eq:quad_regression}) models in logarithmic space, respectively. For each model, thick lines indicate the median relation, while thin lines show the corresponding 16th and 84th percentile regressions. The best-fitting parameters are reported in Table~\ref{Tab:Scalings}.}
    \label{Fig:scalings}
\end{figure*}

The radial profiles presented above show that the thermodynamic state of the filament gas varies systematically with line-mass. We now quantify these trends by deriving scaling relations between $\Lambda$ and the characteristic gas properties of individual filaments.

For each filament, we compute the volume-weighted average of the gas quantity $Q=T,P_e,n_e,Z,K$ within $r_K$ as
\begin{equation}\label{Eq:Qv}
    \langle Q\rangle =
    \frac{\displaystyle\sum_{r\leq r_K} Q(r)\,V(r)}
    {\displaystyle\sum_{r\leq r_K} V(r)},
\end{equation}
where $Q(r)$ is the radial profile shown in Fig.~\ref{Fig:THERMOpro} and defined in Eq.~\ref{Eq:Qprofile}, and $V(r)$ is the total volume of the gas cells that contribute to each radial bin. Since the profiles themselves are volume weighted, Eq.~\ref{Eq:Qv} is equivalent to computing the volume-weighted mean over all gas cells enclosed within $r_K$. For temperature, we additionally compute the mass-weighted average, $\langle T_m\rangle$, using the corresponding $T_m$ profile presented in Fig.~\ref{Fig:THERMOpro} and weighting each radial bin by its enclosed gas mass.\\

The resulting relations are shown in Fig.~\ref{Fig:scalings}, where each blue point represents an individual filament, and the black dashed curves show the median trends. Temperature exhibits an approximately linear relation in logarithmic space. We therefore fit the individual filaments using median quantile regression and adopt
\begin{equation}\label{Eq:lin_regression}
    \log_{10}(T) = a + b\,\log_{10}(x),
\end{equation}
where $x=\Lambda/\Lambda_{\rm med}$ and $\Lambda_{\rm med}=10^{11.67}\,\mathrm{M_\odot\,Mpc^{-1}}$ is the median line-mass of our sample (Sect.~\ref{SubSect:StructuralCorr}). This corresponds to a power-law relation between temperature and filament line-mass.
For the remaining gas properties, the relations show stronger curvature and cannot be adequately described by a single power law over the full line-mass range. We therefore adopt
\begin{equation}\label{Eq:quad_regression}
    \log_{10}(Q) =
    a+b\,\log_{10}(x)
    +c \, \left[\log_{10}(x)\right]^2,
\end{equation}
where $Q=P_e,n_e,Z,K$. The resulting median regression curves are shown in Fig.~\ref{Fig:scalings}, together with the corresponding 16th and 84th percentile regressions. The best-fitting parameters are reported in Table~\ref{Tab:Scalings}.

We find a tight temperature--line-mass relation that extends over almost two orders of magnitude in temperature across the filament population. Both volume and mass-weighted temperatures follow similar trends, with $\langle T \rangle \propto \Lambda^{0.72}$ and $\langle T_m \rangle \propto \Lambda^{0.66}$ (Table~\ref{Tab:Scalings}).
This strong temperature--line-mass relation has a simple physical motivation. 
Following \cite{Lu2024_fils}, for a self-gravitating filament, the characteristic velocity scale is set by its total line-mass, $v \sim (\mathrm{G} \,\Lambda_{\rm tot})^{1/2}$, where G is the gravitational constant. This implies a characteristic temperature $T \propto \mathrm{G} \, \Lambda_{\rm tot}$. Therefore, gas accreting onto the filament can be shock-heated to progressively higher temperatures as the filament line-mass increases \citep{Lu2024_fils}. 
Although the $\Lambda$ measured here corresponds to the gas component rather than the total mass, and the gas-to-total mass ratio may itself vary with $\Lambda$, the observed trend is qualitatively consistent with this picture\footnote{A direct test of this theoretical picture would require measuring the total (gas + dark matter) line-mass, which we defer to future work}. 
Gas accretion is not the only process that can increase the filament temperature. Filament contraction can also compress and heat the gas as the structure evolves. Distinguishing the relative contribution of different heating pathways would require a dedicated analysis of filament evolution and gas flows across multiple snapshots, which we defer to future work.

Compared to previous studies, \citet{GhellerVazza2015,GhellerVazza2016} also found that the mean gas temperature increases with the total filament mass and interpreted this relation as a consequence of the increasing importance of self-gravity in more massive structures. Using a larger suite of simulations including different baryonic physics models, \citet{GhellerVazza2019_surveyTandNTprops_fils} found that, at $z=0$, the mass-weighted temperature scales with the total baryonic mass of the filament, $M_{\rm BM}$, following $T\propto M_{\rm BM}^{0.30-0.37}$. Combining this relation with their scaling  between length and total mass ($L\propto M_{\rm BM}^{0.5}$), we deduce $T\propto (M_{\rm BM}/L)^{0.6-0.74}$. This is remarkably consistent with our findings, despite differences in filament definitions, simulations, and methodologies. 
A similar scaling was already present in our previous analysis of the same TNG300 filament catalogue. \citet{GalarragaEspinosa2022} found that the temperature and gas density of short and long filaments follow $T\propto\rho_{\rm gas}^{0.70}$. Although $\Lambda$ and $\rho_{\rm gas}$ are closely related, our previous work adopted a fixed radius of $0.7$~Mpc whereas here we integrate within the physically motivated radius $r_K$. The present study also uses the full gas-cell dataset rather than a subsampled catalogue, allowing us to extend the analysis to a significantly larger number of individual filaments ($\sim 1000$ vs $5637$). Together with the independent results of \cite{GhellerVazza2015,GhellerVazza2016} and \cite{GhellerVazza2019_surveyTandNTprops_fils}, this agreement suggests that a temperature scaling close to $\langle T\rangle\propto\Lambda^{0.72}$ is a robust feature of cosmic filaments.\\

Pressure and electron density also increase strongly with line-mass, with a clear steepening towards the highest $\Lambda$ values. The pressure relation increases particularly rapidly because both density and temperature increase with $\Lambda$. Interestingly, metallicity also increases with $\Lambda$, but flattens at the high-$\Lambda$ end around $\sim0.1 \,Z_\odot$. We stress that the gas associated with massive haloes and nodes has been masked from the analysis (Sect.~\ref{SubSect:filament_Cat}), so this trend reflects enrichment of the filament gas itself and not contamination from the hot metal-rich halo gas. Entropy rises across the line-mass range probed here, although the median relation shows a slight downturn at the highest line-masses. This may partly reflect the steepening of the electron-density relation, which counteracts the increase in temperature.
However, given the small number of filaments in this regime (see blue points), the apparent downturn is likely to result from sample variance.

These scaling relations show that $\Lambda$ systematically organises the physical state of filament gas, with temperature following the simplest relation and the other quantities showing additional non-linear behaviour. Combined with the morphological trends and the $\Lambda$-dependent thermal structure discussed above, our results show that line-mass is an effective parameter for characterising the diversity of individual cosmic filaments.\\

The structural correlations, thermodynamic profiles, $T_m/T_v$ transition, and scaling relations established above jointly characterise the structure and thermodynamic state of filaments as a function of line-mass $\Lambda$. We now ask whether $\Lambda$ itself can be inferred from observational tracers, without direct gas detections, opening a direct path to applying this framework to current and forthcoming surveys.

\begin{table}
	\centering

    \caption{\justifying \noindent \textup{\textbf{Scaling-relation fit results}. For each relation, we report the best-fitting coefficients, with the corresponding $95\%$ confidence intervals given in brackets. The temperature relations are fitted using the power-law model of Eq.~\ref{Eq:lin_regression}, while the remaining gas properties are fitted using the quadratic model of Eq.~\ref{Eq:quad_regression}.}}
	\label{Tab:Scalings}
	\begin{tabular}{lccr}
		\hline
		$Q$ & a & b & c \\
		\hline
		$\langle T \rangle$ & 5.79 [5.79, 5.8] & 0.72 [0.7, 0.74] & - \\
		$\langle T_m \rangle$ & 5.97 [5.97, 5.98] & 0.66 [0.64, 0.68] & - \\
		$\langle P_e \rangle$ & -7.04 [-7.05, -7.03] & 1.09 [1.06, 1.12] & 0.47 [0.42, 0.53] \\
        $\langle n_e \rangle$ & -5.94 [-5.94, -5.93] & 0.45 [0.44, 0.47] & 0.29 [0.26, 0.33] \\
        $\langle Z \rangle$ & -1.25 [-1.26, -1.24] & 0.33 [0.32, 0.35] & -0.18 [-0.21, -0.14] \\
        $\langle K \rangle$ & 3.12 [3.11, 3.13] & 0.53 [0.51, 0.55] & 0.13 [0.08, 0.17] \\
		\hline
	\end{tabular}
\end{table}

\section{Observational Diagnostics of Cosmic Filaments}
\label{Sect:Observables}

Having established $\Lambda$ as a unifying physical descriptor of cosmic filaments, we now ask how it can be measured observationally. We identify two complementary routes: galaxy richness per unit length, accessible from spectroscopic surveys, and the integrated Compton-$y$ signal, accessible from CMB experiments. Together these define an observational programme for characterising filaments across the cosmic web, which we develop further in our follow-up companion paper.

\subsection{Galaxy-based diagnostics: richness as a proxy for $\Lambda$}
\label{SubSect:GalaxyProxy}

\begin{figure}
    \centering
    \includegraphics[width=1\linewidth]
    {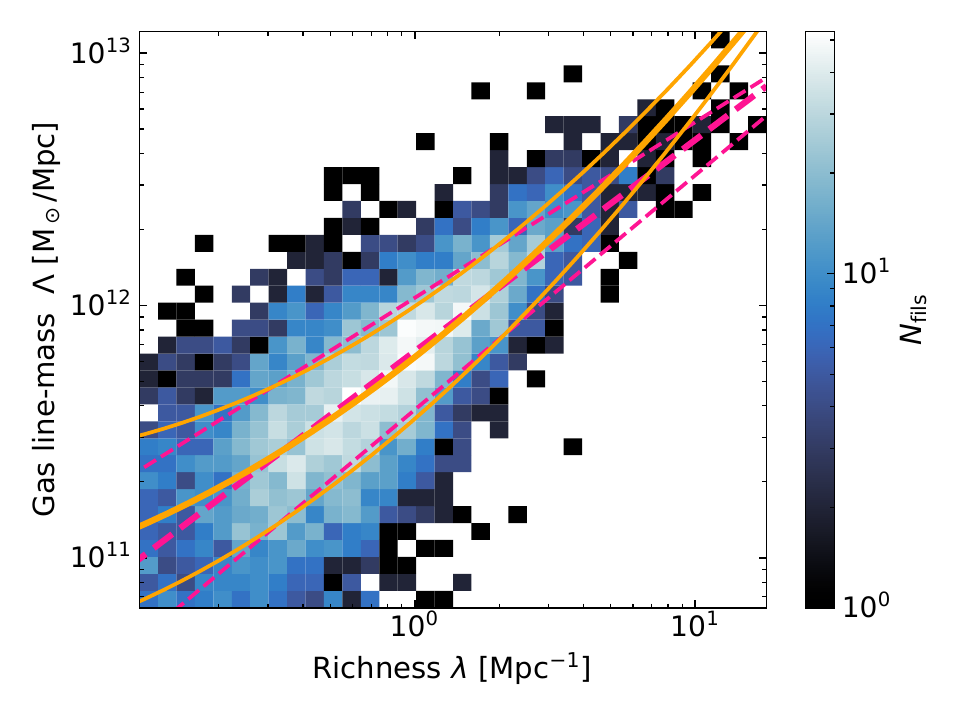}
    \caption{Filament richness per unit length 
    $\lambda$ versus gaseous line-mass $\Lambda$. 
    The colour scale shows the number of filaments 
    in each bin. Pink dashed and orange solid curves 
    show the linear and quadratic quantile-regression 
    fits (Eqs.~\ref{Eq:lin_regression} and 
    \ref{Eq:quad_regression}), respectively. For 
    each model, thick lines indicate the median 
    relation and thin lines show the 16th and 84th 
    percentile regressions. The tight correlation 
    (Pearson coefficient $= 0.80$) demonstrates 
    that $\Lambda$ can be reliably inferred from 
    the galaxy distribution alone. Best-fitting 
    parameters are reported in 
    Table~\ref{Tab:Scaling_Rich_Mass}.}
    \label{Fig:Rich_vs_Lambda}
\end{figure}

\begin{table}
    \centering
    \caption{\justifying \noindent \textup{\textbf{Best-fitting coefficients} for the 
    scaling relation between filament richness 
    $\lambda$ and gaseous line-mass $\Lambda$ 
    (Fig.~\ref{Fig:Rich_vs_Lambda}), with $95\%$ 
    confidence intervals in brackets. The linear 
    model uses Eq.~\ref{Eq:lin_regression} and 
    the quadratic model uses 
    Eq.~\ref{Eq:quad_regression}.}}
    \label{Tab:Scaling_Rich_Mass}
    \begin{tabular}{lccr}
        \hline
         & $a$ & $b$ & $c$ \\
        \hline
        Lin. &
        11.82 [11.81, 11.83] &
        0.84 [0.82, 0.86] &
        -- \\
        Quad. &
        11.79 [11.79, 11.80] &
        0.88 [0.85, 0.90] &
        0.20 [0.16, 0.24] \\
        \hline
    \end{tabular}
\end{table}

The scaling relations derived in Sect.~\ref{Sect:FilProperties} connect $\Lambda$ to the full thermodynamic state of filament gas. 
Here we ask whether $\Lambda$ can be inferred directly from the galaxy distribution, without direct access to gas observations, opening a route to applying this framework to current and forthcoming spectroscopic surveys including SDSS, DESI, Subaru PFS, and Euclid.

Galaxy number-density profiles of filaments, computed following \citet{GalarragaEspinosa2020} but binned by $\Lambda$ rather than length, show a clear and systematic ordering: 
high line-mass filaments host larger galaxy overdensities around their spines across the full radial range probed. This ordering mirrors the thermodynamic sequence established in Sect.~\ref{SubSect:ThermoPro}, showing that the same $\Lambda$ that organises the gas properties also organises the galaxy-density content. The number-density profiles can be seen in Appendix~\ref{APP:ngal}.\\

We define for each filament the richness per unit length as
\begin{equation}\label{Eq:richness}
    \lambda = 
    \frac{N_\mathrm{gal}(r < 1\,\mathrm{Mpc})}
    {L_\mathrm{fil}},
\end{equation}
where $N_\mathrm{gal}$ is the number of galaxies within $1\,\mathrm{Mpc}$ of the filament spine. This definition is inspired by the use of (optical) richness as a mass proxy in galaxy clusters; the aperture of $1\,\mathrm{Mpc}$ is motivated by the typical filament radius inferred from galaxy number-density profiles \citep[e.g.,][]{Wang2024fil_radius}.

Figure~\ref{Fig:Rich_vs_Lambda} shows the resulting $\lambda$--$\Lambda$ relation. We find a tight correlation with a Pearson coefficient of $0.80$, with modest scatter towards the low-richness end where the discrete galaxy distribution provides a poorer sampling of the filament. We fit the relation using the linear and quadratic models of Eqs.~\ref{Eq:lin_regression} and \ref{Eq:quad_regression}, finding best-fit parameters reported in  Table~\ref{Tab:Scaling_Rich_Mass}. The linear model gives $\log_{10}(\Lambda) = 11.82 + 0.84\,\log_{10}(\lambda)$ and provides a good description over most of the sampled range. However, the non-zero quadratic coefficient ($c = 0.20$) better captures a mild curvature at the low-richness end.
For both models, we measure a RMS scatter of around $0.26$~dex.

This tight correlation is particularly encouraging from an observational perspective. For a filament with a known galaxy count and length, $\Lambda$ can be estimated to within the scatter of this relation without any gas observations. The proxy is immediately applicable to existing SDSS filament catalogues  \citep[e.g.][]{Tempel2016bisous, Malavasi2020_sdss}, exploitable with ongoing DESI spectroscopy \citep{DESICollaboration2016}, and will benefit from the deeper and wider samples provided by Subaru-PFS and Euclid within the next few years.
We stress that the present calibration applies to the prominent, well-traced filament population selected in Sect.~\ref{SubSect:ResolutionCut}, and depends on both the tracer-galaxy selection and the adopted aperture\footnote{In particular, observers applying this relation should adopt a minimum stellar mass threshold close to $M_* > 10^9\,\Msun{}$ and radial aperture $R = 1 \,\mathrm{Mpc}$ used here when constructing the richness estimator from spectroscopic catalogues.}.

\subsection{Thermal Sunyaev--Zel'dovich diagnostics: $Y^{\rm int}$ as a probe of filament thermodynamics}\label{SubSect:Yint}

\begin{figure}
    \centering
    \includegraphics[width=1\linewidth]{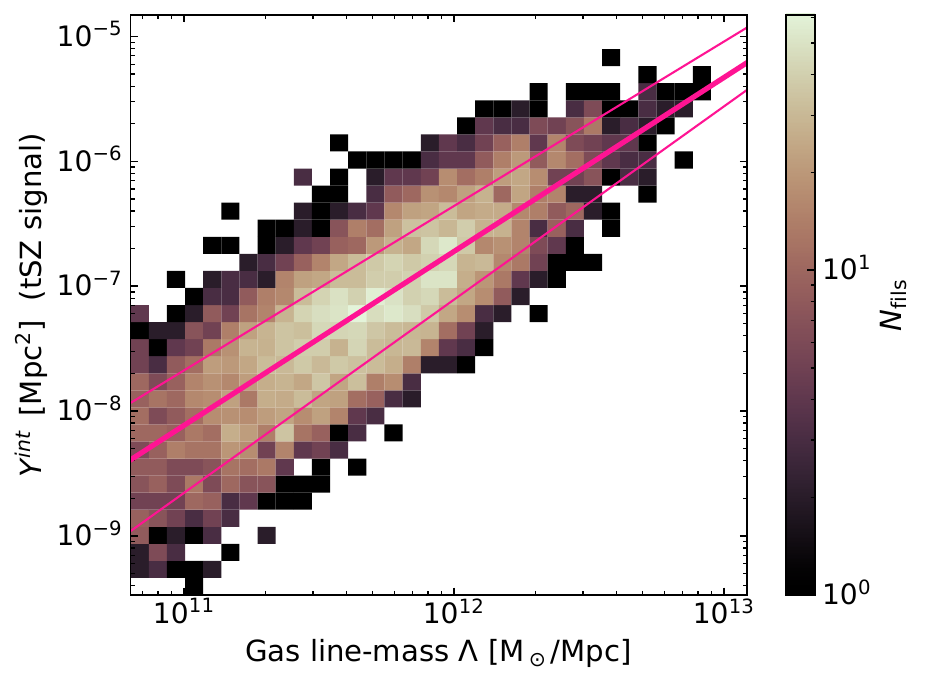}
    \caption{$Y^{\rm int}$--$\Lambda$ scaling relation. The relation is well described by a power-law model, $\log_{10}(Y^{\rm int}) = -7.19 + 1.39\,\log_{10}(x)$, shown by the thick pink line, where $x=\Lambda/\Lambda_{\rm med}$. Thin pink lines indicate the corresponding 16th and 84th percentile regressions.}
    \label{Fig:Yint}
\end{figure}

A complementary observational probe of filament gas is provided by the thermal Sunyaev-Zel'dovich (tSZ) effect, which traces the integrated electron pressure along the line of sight and is particularly sensitive to the diffuse warm-hot gas that is difficult to detect in X-rays \citep{DeGraaff2019, Tanimura2019}. The Compton-$y$ parameter at a given position on the sky is defined as
\begin{equation}
y = \frac{\sigma_{\rm T}}{m_e c^2}
\int P_e\,{\rm d}l ,
\end{equation}
where $\sigma_{\rm T}$ is the Thomson cross-section, $m_e$ is the electron mass, and $c$ is the speed of light. The integrated tSZ signal, $Y_{\rm SZ}=\int y\,{\rm d}\Omega$, is obtained by further integrating the Compton-$y$ signal over the projected angular extent of the structure, therefore measuring the structure's total thermal-pressure content. Here, we consider the corresponding intrinsic integrated Compton-$y$ parameter,
\begin{equation}
    Y^{\rm int} \equiv D_A^2 \, Y_{\rm SZ} = \frac{\sigma_{\rm T}}{m_e c^2}
\int_V P_e\,{\rm d}V ,
\end{equation}
where $D_A$ is the angular-diameter distance. The term ``intrinsic'' indicates that the geometrical $D_A^{-2}$ dependence of the observed angular signal has been removed, such that $Y^{\rm int}$ directly characterises the volume-integrated thermal pressure of the gas
\citep[e.g.,][]{Nagai2006}. Considering cylindrical geometry, the intrinsic integrated Compton-$y$ parameter for a filament becomes:
\begin{equation}
\label{Eq:Yint}
Y^{\rm int} = \frac{\sigma_{\rm T}}{m_e c^2} 
\, 2\pi \, L_{\rm fil} 
\int_0^{r_{\rm fil}} P_e(r) \, r \, {\rm d}r,
\end{equation}
where $P_e(r)$ is the radial electron pressure profile and $r_{\rm fil} = r_K$.\\

Figure~\ref{Fig:Yint} shows the resulting $Y^{\rm int}$--$\Lambda$ scaling relation, directly obtained by integrating the electron pressure profiles following Eq~\ref{Eq:Yint}. We see that $Y^{\rm int}$ increases systematically with $\Lambda$, reflecting the combined increase in electron pressure and filament radius with line-mass. The relation is well described by a power-law model, $\log_{10}(Y^{\rm int}) = -7.19 + 1.39 \,\log_{10}(x)$, where $x = \Lambda/\Lambda_{\rm med}$, with an RMS scatter of $0.4$~dex. The fit parameters' $95\%$ confidence intervals are [-7.2, -7.17] and [1.36, 1.43], respectively. \\

Together, the optical richness proxy $\lambda$ and the tSZ $Y^{\rm int}$ provide two independent observational routes to the thermodynamic characterisation of cosmic filaments. 
We note that, although the total integrated signal $Y_{SZ}$ does not depend on orientation when the whole filament is enclosed, its Compton-$y$ morphology does: filaments aligned with the line of sight subtend a smaller solid angle but reach higher $y$, whereas filaments lying in the plane of the sky are more extended with lower $y$. Aperture-based and stacking analyses that assume random orientations can therefore be affected by a systematic geometric bias. A companion paper will present a comprehensive study of both the thermal and kinetic Sunyaev-Zel'dovich signals from cosmic filaments as a function of $\Lambda$ and viewing angle, with the goal of constraining thermal and non-thermal pressure profiles across the filament population. This will include modelling of non-thermal pressure support as a function of distance from the 
filament spine and the development of observational analysis strategies to exploit current and forthcoming tSZ and kSZ data from \textit{Planck}, ACT, and the Simons Observatory.

\section{Conclusions}\label{Sect:Conclusions}

We have used the gaseous line-mass $\Lambda \equiv M_{\rm gas}/L_{\rm fil}$, measured within the entropy radius $r_K$, as an organising parameter for the physical diversity of cosmic filaments. Using IllustrisTNG300-1 at $z = 0$, we have characterised the structural, thermodynamic, and chemical properties of $5637$ filaments spanning more than two decades in $\Lambda$, and demonstrated that it organises filament gas into a continuous, well-ordered sequence analogous to the role of mass in galaxy cluster physics. Our main conclusions are as follows.

\begin{enumerate}

  \item \textbf{$\Lambda$ organises filament diversity.} Filaments form a continuous sequence with $\Lambda$: from long, thin, cool, and metal-poor structures at low $\Lambda$ to shorter, wider, hotter, and more enriched ones at high $\Lambda$. Filament length alone is a poor predictor of 
  thermodynamic state; $\Lambda$ captures this diversity more effectively (Sect.~\ref{SubSect:StructuralCorr}, Fig.~\ref{Fig:binnedstatistics}).

  \item \textbf{Thermodynamic profiles are ordered by $\Lambda$.} Radial profiles of temperature, electron density, pressure, entropy, and metallicity are all clearly ordered with $\Lambda$ (Sect.~\ref{Sect:FilProperties}, Fig.~\ref{Fig:THERMOpro}). Core temperatures span nearly two orders of magnitude, from a few $10^5\,\mathrm{K}$ at low $\Lambda$ to ${\sim}\,10^7\,\mathrm{K}$ at high $\Lambda$, spanning the full WHIM temperature window and connecting the high-$\Lambda$ end to the thermodynamic regime of cluster outskirts. 
  The entropy enhancement around $r_K$ becomes increasingly pronounced with $\Lambda$, revealing a stronger thermodynamic contrast between the inner filament gas and the surrounding medium.

  \item \textbf{Scaling relations connect $\Lambda$ to all thermodynamic quantities.} Well-defined scaling relations emerge between $\Lambda$ and volume-weighted temperature, mass-weighted temperature, electron pressure, electron density, entropy, and metallicity (Table~\ref{Tab:Scalings}, Fig.~\ref{Fig:scalings}). Temperature follows a simple power law, $\langle T\rangle \propto \Lambda^{0.72}$. This is consistent with the expectation that the characteristic temperature of self-gravitating filaments increases with their total line-mass, allowing accreting gas to be shock-heated to higher temperatures in deeper gravitational potentials \citep{Lu2024_fils}. Other heating pathways, including gas compression associated with filament contraction, may also contribute. The agreement with independent results from different simulation suites and filament finders \citep{GhellerVazza2019_surveyTandNTprops_fils,Lu2024_fils} supports a consistent connection between filament temperature and line-mass. The remaining quantities show additional curvature at high $\Lambda$ and are well described by quadratic models in logarithmic space.

  \item \textbf{Thermal structure transitions with $\Lambda$.} 
  The ratio between mass- and volume-weighted temperatures, $T_m/T_v$, decreases from $\sim2$ in the interiors of low-$\Lambda$ filaments, where the hotter gas is preferentially denser, to values below unity near $r_K$ in high-$\Lambda$ filaments, where the hotter gas is preferentially diffuse (Fig.~\ref{Fig:TmTv}). This reveals a systematic change in how thermal energy is distributed within filaments. Since X-ray emission preferentially traces dense gas, while the tSZ signal traces integrated thermal pressure, this transition is expected to modify their relative sensitivity to filament gas as a function of $\Lambda$.
  
  \item \textbf{Metallicity increases systematically with $\Lambda$.} Filament gas metallicity rises from $Z \sim 0.03\,Z_\odot$ at low $\Lambda$ to $Z \sim 0.1$--$0.13\,Z_\odot$ at high $\Lambda$ (Fig.~\ref{Fig:THERMOpro}), consistent with X-ray measurements of individual massive filaments 
  \citep{Veronica2024, Dietl2024}. This enrichment is present in the diffuse filament gas after masking cluster and node contributions, supporting large-scale metal transport driven by stellar and AGN feedback within the TNG model.

  \item \textbf{First measurement of the gaseous line-mass function.} We present the first measurement of the filament gas line-mass function from a cosmological hydrodynamical simulation (Fig.~\ref{Fig:Lambda_fils}). The distribution is best described by a Tinker-like functional form ($\chi^2_\nu = 1.09$), with a broad  low-$\Lambda$ regime and a steep exponential decline at high $\Lambda$, mirroring the suppression of the halo mass function at high masses. This provides a statistical baseline and fitting formula for future observational surveys and, combined with the richness proxy calibration, a testable prediction for the distribution of galaxy richness per unit length in current SDSS and forthcoming DESI filament catalogues.

  \item \textbf{Galaxy richness is a robust observational proxy for $\Lambda$.} The galaxy richness per unit length $\lambda$ correlates tightly with $\Lambda$ (Pearson $r = 0.80$; Fig.~\ref{Fig:Rich_vs_Lambda}, Table~\ref{Tab:Scaling_Rich_Mass}), allowing the thermodynamic state of a filament to be inferred from the galaxy distribution alone, without gas observations. This opens a direct path to applications with spectroscopic surveys, such as SDSS, DESI, Subaru-PFS, and Euclid.

  \item \textbf{$Y^{\rm int}$ provides a complementary tSZ proxy for $\Lambda$.} 
  The integrated Compton-$y$ parameter $Y^{\rm int}$ scales systematically with 
  $\Lambda$ (Fig.~\ref{Fig:Yint}), providing an independent observational route to filament thermodynamic characterisation via the tSZ effect. 
\end{enumerate}

The temperature sequence identified here ---spanning nearly two orders of magnitude from a few $10^5\,\mathrm{K}$ at low $\Lambda$ to nearly $10^7\,\mathrm{K}$ at high $\Lambda$--- directly connects to the missing baryon problem at low redshift. Low-$\Lambda$ filaments host the cooler, more diffuse WHIM that is most challenging to detect observationally, while high-$\Lambda$ filaments approach the thermodynamic conditions of cluster outskirts where tSZ and X-ray detections are already within reach \citep[e.g.,][]{DeGraaff2019, Tanimura2019, Veronica2024}. The $\Lambda$ framework therefore not only organises filament physics but directly guides the observational strategy for completing the baryon census.

The results presented in this work establish $\Lambda$ as an effective organising parameter for the physical diversity of cosmic filaments, analogous to the role of mass for dark matter haloes. The scaling relations, line-mass function, richness proxy, and $Y^{\rm int}$--$\Lambda$ relation derived here provide a physically motivated framework for the classification, comparison, and observational characterisation of filaments across the large-scale web.

The next step is to connect these results directly to observations. The richness proxy calibration enables immediate application to spectroscopic galaxy surveys. The $Y^{\rm int}$--$\Lambda$ relation provides a complementary tSZ route accessible to \textit{Planck}, ACT, and the Simons Observatory; a companion paper will extend this to a joint tSZ and kSZ analysis aimed at constraining both thermal and non-thermal pressure profiles across the filament population as a function of $\Lambda$ and viewing angle. High-resolution X-ray spectroscopy with XRISM may provide complementary constraints on the temperature, chemical enrichment, and gas dynamics of the high-$\Lambda$ filaments.
On longer timescales, future X-ray facilities 
such as NewAthena may enable entropy profiles to be reconstructed for individual filaments over a broad range of $\Lambda$, providing a direct observational test of the thermodynamic framework developed here.

\section*{Acknowledgements}

D.G.-E. thanks Nir Mandelker for useful discussions during the preliminary stages of this work. D.G.-E. also thanks Massimiliano Bonamente for discussions that inspired the richness--line-mass scaling relation.
This work made use of outputs from the ByoPiC project (\url{https://byopic.eu}), funded by the European Research Council (ERC) under the European Union's Horizon 2020 research and innovation programme (grant agreement ERC-2015-AdG 695561). This work is supported by the National Science Foundation (NSF) grant AST 2511137 and the Yale Center for Research Computing facilities and staff. I.M. acknowledges support from the Dean's Emerging Scholars Research Award from the Yale Graduate School of Arts \& Sciences. K.N. acknowledges support from JSPS KAKENHI grants 24H00002, 24H00241, JP25K01032, and the  International Leading Research (ILR) project, JP22K21349. K.N. also acknowledges support from the Kavli IPMU, the World Premier Research Center Initiative (WPI), UTIAS, and the University of Tokyo. ChatGPT-5 (OpenAI, 2025) was used to assist with the formatting of selected tables and to improve the clarity of the language. All AI-assisted content was carefully reviewed, edited, and verified for accuracy by the authors, who take full responsibility for the final manuscript.
\textbf{Author contributions:}
We describe the different author contributions using
the CRediT\footnote{\url{https://credit.niso.org/}} (Contributor Roles Taxonomy) system.
    \textbf{D.G.-E.}: conceptualization; data curation; formal analysis;
    investigation; methodology; resources; software; validation; visualization; writing - original draft, writing – review \& editing.
    \textbf{D.N.}: conceptualization; investigation; methodology; supervision; writing - original draft; writing – review \& editing.
    \textbf{I.M.}: conceptualization; methodology; writing – review \& editing.
    \textbf{M.L.}: conceptualization; writing – review \& editing.
    \textbf{N.A.}: writing – review \& editing.
    \textbf{K.N.}: writing – review \& editing.

\section*{Data Availability}

The IllustrisTNG simulations are publicly 
available at \url{https://www.tng-project.org} 
\citep{Nelson2019_TNGdata_release}. The filament 
catalogue and derived gas property profiles 
underlying this article will be shared on 
reasonable request to the corresponding author.

\bibliographystyle{mnras}
\bibliography{main} 

@ARTICLE{Inutsuka1992,
       author = {{Inutsuka}, Shu-Ichiro and {Miyama}, Shoken M.},
        title = "{Self-similar Solutions and the Stability of Collapsing Isothermal Filaments}",
      journal = {\apj},
         year = 1992,
        month = apr,
       volume = {388},
        pages = {392},
          doi = {10.1086/171162},
       adsurl = {https://ui.adsabs.harvard.edu/abs/1992ApJ...388..392I}
}

@ARTICLE{DESICollaboration2016,
       author = {{DESI Collaboration} and {Aghamousa}, Amir and {Aguilar}, Jessica and {Ahlen}, Steve and {Alam}, Shadab and {Allen}, Lori E. and {Allende Prieto}, Carlos and {Annis}, James and {Bailey}, Stephen and {Balland}, Christophe and {Ballester}, Otger and {Baltay}, Charles and {Beaufore}, Lucas and {Bebek}, Chris and {Beers}, Timothy C. and {Bell}, Eric F. and {Bernal}, Jos{\'e} Luis and {Besuner}, Robert and {Beutler}, Florian and {Blake}, Chris and {Bleuler}, Hannes and {Blomqvist}, Michael and {Blum}, Robert and {Bolton}, Adam S. and {Briceno}, Cesar and {Brooks}, David and {Brownstein}, Joel R. and {Buckley-Geer}, Elizabeth and {Burden}, Angela and {Burtin}, Etienne and {Busca}, Nicolas G. and {Cahn}, Robert N. and {Cai}, Yan-Chuan and {Cardiel-Sas}, Laia and {Carlberg}, Raymond G. and {Carton}, Pierre-Henri and {Casas}, Ricard and {Castander}, Francisco J. and {Cervantes-Cota}, Jorge L. and {Claybaugh}, Todd M. and {Close}, Madeline and {Coker}, Carl T. and {Cole}, Shaun and {Comparat}, Johan and {Cooper}, Andrew P. and {Cousinou}, M.-C. and {Crocce}, Martin and {Cuby}, Jean-Gabriel and {Cunningham}, Daniel P. and {Davis}, Tamara M. and {Dawson}, Kyle S. and {de la Macorra}, Axel and {De Vicente}, Juan and {Delubac}, Timoth{\'e}e and {Derwent}, Mark and {Dey}, Arjun and {Dhungana}, Govinda and {Ding}, Zhejie and {Doel}, Peter and {Duan}, Yutong T. and {Ealet}, Anne and {Edelstein}, Jerry and {Eftekharzadeh}, Sarah and {Eisenstein}, Daniel J. and {Elliott}, Ann and {Escoffier}, St{\'e}phanie and {Evatt}, Matthew and {Fagrelius}, Parker and {Fan}, Xiaohui and {Fanning}, Kevin and {Farahi}, Arya and {Farihi}, Jay and {Favole}, Ginevra and {Feng}, Yu and {Fernandez}, Enrique and {Findlay}, Joseph R. and {Finkbeiner}, Douglas P. and {Fitzpatrick}, Michael J. and {Flaugher}, Brenna and {Flender}, Samuel and {Font-Ribera}, Andreu and {Forero-Romero}, Jaime E. and {Fosalba}, Pablo and {Frenk}, Carlos S. and {Fumagalli}, Michele and {Gaensicke}, Boris T. and {Gallo}, Giuseppe and {Garcia-Bellido}, Juan and {Gaztanaga}, Enrique and {Pietro Gentile Fusillo}, Nicola and {Gerard}, Terry and {Gershkovich}, Irena and {Giannantonio}, Tommaso and {Gillet}, Denis and {Gonzalez-de-Rivera}, Guillermo and {Gonzalez-Perez}, Violeta and {Gott}, Shelby and {Graur}, Or and {Gutierrez}, Gaston and {Guy}, Julien and {Habib}, Salman and {Heetderks}, Henry and {Heetderks}, Ian and {Heitmann}, Katrin and {Hellwing}, Wojciech A. and {Herrera}, David A. and {Ho}, Shirley and {Holland}, Stephen and {Honscheid}, Klaus and {Huff}, Eric and {Hutchinson}, Timothy A. and {Huterer}, Dragan and {Hwang}, Ho Seong and {Illa Laguna}, Joseph Maria and {Ishikawa}, Yuzo and {Jacobs}, Dianna and {Jeffrey}, Niall and {Jelinsky}, Patrick and {Jennings}, Elise and {Jiang}, Linhua and {Jimenez}, Jorge and {Johnson}, Jennifer and {Joyce}, Richard and {Jullo}, Eric and {Juneau}, St{\'e}phanie and {Kama}, Sami and {Karcher}, Armin and {Karkar}, Sonia and {Kehoe}, Robert and {Kennamer}, Noble and {Kent}, Stephen and {Kilbinger}, Martin and {Kim}, Alex G. and {Kirkby}, David and {Kisner}, Theodore and {Kitanidis}, Ellie and {Kneib}, Jean-Paul and {Koposov}, Sergey and {Kovacs}, Eve and {Koyama}, Kazuya and {Kremin}, Anthony and {Kron}, Richard and {Kronig}, Luzius and {Kueter-Young}, Andrea and {Lacey}, Cedric G. and {Lafever}, Robin and {Lahav}, Ofer and {Lambert}, Andrew and {Lampton}, Michael and {Landriau}, Martin and {Lang}, Dustin and {Lauer}, Tod R. and {Le Goff}, Jean-Marc and {Le Guillou}, Laurent and {Le Van Suu}, Auguste and {Lee}, Jae Hyeon and {Lee}, Su-Jeong and {Leitner}, Daniela and {Lesser}, Michael and {Levi}, Michael E. and {L'Huillier}, Benjamin and {Li}, Baojiu and {Liang}, Ming and {Lin}, Huan and {Linder}, Eric and {Loebman}, Sarah R. and {Luki{\'c}}, Zarija and {Ma}, Jun and {MacCrann}, Niall and {Magneville}, Christophe and {Makarem}, Laleh and {Manera}, Marc and {Manser}, Christopher J. and {Marshall}, Robert and {Martini}, Paul and {Massey}, Richard and {Matheson}, Thomas and {McCauley}, Jeremy and {McDonald}, Patrick and {McGreer}, Ian D. and {Meisner}, Aaron and {Metcalfe}, Nigel and {Miller}, Timothy N. and {Miquel}, Ramon and {Moustakas}, John and {Myers}, Adam and {Naik}, Milind and {Newman}, Jeffrey A. and {Nichol}, Robert C. and {Nicola}, Andrina and {Nicolati da Costa}, Luiz and {Nie}, Jundan and {Niz}, Gustavo and {Norberg}, Peder and {Nord}, Brian and {Norman}, Dara and {Nugent}, Peter and {O'Brien}, Thomas and {Oh}, Minji and {Olsen}, Knut A.~G.},
        title = "{The DESI Experiment Part I: Science,Targeting, and Survey Design}",
      journal = {arXiv e-prints},
         year = 2016,
        month = oct,
          eid = {arXiv:1611.00036},
        pages = {arXiv:1611.00036},
          doi = {10.48550/arXiv.1611.00036},
archivePrefix = {arXiv},
       eprint = {1611.00036},
 primaryClass = {astro-ph.IM},
       adsurl = {https://ui.adsabs.harvard.edu/abs/2016arXiv161100036D}
}

@ARTICLE{Asplund2009,
       author = {{Asplund}, Martin and {Grevesse}, Nicolas and {Sauval}, A. Jacques and {Scott}, Pat},
        title = "{The Chemical Composition of the Sun}",
      journal = {\araa},
         year = 2009,
        month = sep,
       volume = {47},
       number = {1},
        pages = {481-522},
          doi = {10.1146/annurev.astro.46.060407.145222},
archivePrefix = {arXiv},
       eprint = {0909.0948},
 primaryClass = {astro-ph.SR},
       adsurl = {https://ui.adsabs.harvard.edu/abs/2009ARA&A..47..481A}
}

@ARTICLE{GalarragaEspinosa2022,
       author = {{Gal{\'a}rraga-Espinosa}, Daniela and {Langer}, Mathieu and {Aghanim}, Nabila},
        title = "{Relative distribution of dark matter, gas, and stars around cosmic filaments in the IllustrisTNG simulation}",
      journal = {\aap},
         year = 2022,
        month = may,
       volume = {661},
          eid = {A115},
        pages = {A115},
          doi = {10.1051/0004-6361/202141974},
archivePrefix = {arXiv},
       eprint = {2109.06198},
 primaryClass = {astro-ph.CO},
       adsurl = {https://ui.adsabs.harvard.edu/abs/2022A&A...661A.115G}
}

@ARTICLE{Borzyszkowski2017_ZOMG1,
       author = {{Borzyszkowski}, Mikolaj and {Porciani}, Cristiano and {Romano-D{\'\i}az}, Emilio and {Garaldi}, Enrico},
        title = "{ZOMG - I. How the cosmic web inhibits halo growth and generates assembly bias}",
      journal = {\mnras},
         year = 2017,
        month = jul,
       volume = {469},
       number = {1},
        pages = {594-611},
          doi = {10.1093/mnras/stx873},
archivePrefix = {arXiv},
       eprint = {1610.04231},
 primaryClass = {astro-ph.CO},
       adsurl = {https://ui.adsabs.harvard.edu/abs/2017MNRAS.469..594B}
}

@ARTICLE{GalarragaEspinosa2024_MTNG,
       author = {{Gal{\'a}rraga-Espinosa}, Daniela and {Cadiou}, Corentin and {Gouin}, C{\'e}line and {White}, Simon D.~M. and {Springel}, Volker and {Pakmor}, R{\"u}diger and {Hadzhiyska}, Boryana and {Bose}, Sownak and {Ferlito}, Fulvio and {Hernquist}, Lars and {Kannan}, Rahul and {Barrera}, Monica and {Maria Delgado}, Ana and {Hern{\'a}ndez-Aguayo}, C{\'e}sar},
        title = "{Evolution of cosmic filaments in the MTNG simulation}",
      journal = {\aap},
         year = 2024,
        month = apr,
       volume = {684},
          eid = {A63},
        pages = {A63},
          doi = {10.1051/0004-6361/202347982},
archivePrefix = {arXiv},
       eprint = {2309.08659},
 primaryClass = {astro-ph.CO},
       adsurl = {https://ui.adsabs.harvard.edu/abs/2024A&A...684A..63G}
}

@ARTICLE{Bahe2025_filaments,
       author = {{Bah{\'e}}, Yannick M. and {Jablonka}, Pascale},
        title = "{Galaxies in the simulated cosmic web: I. Filament identification and properties}",
      journal = {\aap},
         year = 2025,
        month = oct,
       volume = {702},
          eid = {A145},
        pages = {A145},
          doi = {10.1051/0004-6361/202554079},
archivePrefix = {arXiv},
       eprint = {2502.06484},
 primaryClass = {astro-ph.GA},
       adsurl = {https://ui.adsabs.harvard.edu/abs/2025A&A...702A.145B}
}

@ARTICLE{Zhu2021_filaments_evo_z,
       author = {{Zhu}, Weishan and {Zhang}, Fupeng and {Feng}, Long-Long},
        title = "{Profiles of Cosmic Filaments Since z = 4.0 in Cosmological Hydrodynamical Simulation}",
      journal = {\apj},
         year = 2021,
        month = oct,
       volume = {920},
       number = {1},
          eid = {2},
        pages = {2},
          doi = {10.3847/1538-4357/ac15f1},
archivePrefix = {arXiv},
       eprint = {2107.08663},
 primaryClass = {astro-ph.CO},
       adsurl = {https://ui.adsabs.harvard.edu/abs/2021ApJ...920....2Z}
}

@ARTICLE{Veronica2024,
       author = {{Veronica}, Angie and {Reiprich}, Thomas H. and {Pacaud}, Florian and {Ota}, Naomi and {Aschersleben}, Jann and {Biffi}, Veronica and {Bulbul}, Esra and {Clerc}, Nicolas and {Dolag}, Klaus and {Erben}, Thomas and {Gatuzz}, Efrain and {Ghirardini}, Vittorio and {Kerp}, J{\"u}rgen and {Klein}, Matthias and {Liu}, Ang and {Liu}, Teng and {Migkas}, Konstantinos and {Ramos-Ceja}, Miriam E. and {Sanders}, Jeremy and {Spinelli}, Claudia},
        title = "{The eROSITA view of the Abell 3391/95 field. Cluster outskirts and filaments}",
      journal = {\aap},
         year = 2024,
        month = jan,
       volume = {681},
          eid = {A108},
        pages = {A108},
          doi = {10.1051/0004-6361/202347037},
archivePrefix = {arXiv},
       eprint = {2311.07488},
 primaryClass = {astro-ph.CO},
       adsurl = {https://ui.adsabs.harvard.edu/abs/2024A&A...681A.108V}
}

@ARTICLE{Dietl2024,
       author = {{Dietl}, J. and {Pacaud}, F. and {Reiprich}, T.~H. and {Veronica}, A. and {Migkas}, K. and {Spinelli}, C. and {Dolag}, K. and {Seidel}, B.},
        title = "{Discovery of a > 13 Mpc long X-ray filament between two galaxy clusters beyond three times their virial radii}",
      journal = {\aap},
         year = 2024,
        month = nov,
       volume = {691},
          eid = {A286},
        pages = {A286},
          doi = {10.1051/0004-6361/202449354},
archivePrefix = {arXiv},
       eprint = {2401.17281},
 primaryClass = {astro-ph.CO},
       adsurl = {https://ui.adsabs.harvard.edu/abs/2024A&A...691A.286D}
}

@ARTICLE{Wang2024fil_radius,
       author = {{Wang}, Wei and {Wang}, Peng and {Guo}, Hong and {Kang}, Xi and {Libeskind}, Noam I. and {Gal{\'a}rraga-Espinosa}, Daniela and {Springel}, Volker and {Kannan}, Rahul and {Hernquist}, Lars and {Pakmor}, R{\"u}diger and {Yu}, Hao-Ran and {Bose}, Sownak and {Guo}, Quan and {Yu}, Luo and {Hern{\'a}ndez-Aguayo}, C{\'e}sar},
        title = "{The boundary of cosmic filaments}",
      journal = {\mnras},
         year = 2024,
        month = aug,
       volume = {532},
       number = {4},
        pages = {4604-4615},
          doi = {10.1093/mnras/stae1801},
archivePrefix = {arXiv},
       eprint = {2402.11678},
 primaryClass = {astro-ph.CO},
       adsurl = {https://ui.adsabs.harvard.edu/abs/2024MNRAS.532.4604W}
}

@ARTICLE{GaneshaiahVeena2018_CosmicBalletI,
       author = {{Ganeshaiah Veena}, Punyakoti and {Cautun}, Marius and {van de Weygaert}, Rien and {Tempel}, Elmo and {Jones}, Bernard J.~T. and {Rieder}, Steven and {Frenk}, Carlos S.},
        title = "{The Cosmic Ballet: spin and shape alignments of haloes in the cosmic web}",
      journal = {\mnras},
         year = 2018,
        month = nov,
       volume = {481},
       number = {1},
        pages = {414-438},
          doi = {10.1093/mnras/sty2270},
archivePrefix = {arXiv},
       eprint = {1805.00033},
 primaryClass = {astro-ph.CO},
       adsurl = {https://ui.adsabs.harvard.edu/abs/2018MNRAS.481..414G}
}

@ARTICLE{Springel2018_TNG,
       author = {{Springel}, Volker and {Pakmor}, R{\"u}diger and {Pillepich}, Annalisa and {Weinberger}, Rainer and {Nelson}, Dylan and {Hernquist}, Lars and {Vogelsberger}, Mark and {Genel}, Shy and {Torrey}, Paul and {Marinacci}, Federico and {Naiman}, Jill},
        title = "{First results from the IllustrisTNG simulations: matter and galaxy clustering}",
      journal = {\mnras},
         year = 2018,
        month = mar,
       volume = {475},
       number = {1},
        pages = {676-698},
          doi = {10.1093/mnras/stx3304},
archivePrefix = {arXiv},
       eprint = {1707.03397},
 primaryClass = {astro-ph.GA},
       adsurl = {https://ui.adsabs.harvard.edu/abs/2018MNRAS.475..676S}
}

@ARTICLE{Marinacci2018,
       author = {{Marinacci}, Federico and {Vogelsberger}, Mark and {Pakmor}, R{\"u}diger and {Torrey}, Paul and {Springel}, Volker and {Hernquist}, Lars and {Nelson}, Dylan and {Weinberger}, Rainer and {Pillepich}, Annalisa and {Naiman}, Jill and {Genel}, Shy},
        title = "{First results from the IllustrisTNG simulations: radio haloes and magnetic fields}",
      journal = {\mnras},
         year = 2018,
        month = nov,
       volume = {480},
       number = {4},
        pages = {5113-5139},
          doi = {10.1093/mnras/sty2206},
archivePrefix = {arXiv},
       eprint = {1707.03396},
 primaryClass = {astro-ph.CO},
       adsurl = {https://ui.adsabs.harvard.edu/abs/2018MNRAS.480.5113M}
}

@ARTICLE{Mandelker2021,
       author = {{Mandelker}, Nir and {van den Bosch}, Frank C. and {Springel}, Volker and {van de Voort}, Freeke and {Burchett}, Joseph N. and {Butsky}, Iryna S. and {Nagai}, Daisuke and {Oh}, S. Peng},
        title = "{Thermal Instabilities and Shattering in the High-redshift WHIM: Convergence Criteria and Implications for Low-metallicity Strong H I Absorbers}",
      journal = {\apj},
         year = 2021,
        month = dec,
       volume = {923},
       number = {1},
          eid = {115},
        pages = {115},
          doi = {10.3847/1538-4357/ac2d29},
archivePrefix = {arXiv},
       eprint = {2107.03395},
 primaryClass = {astro-ph.CO},
       adsurl = {https://ui.adsabs.harvard.edu/abs/2021ApJ...923..115M}
}

@ARTICLE{Aung2019,
       author = {{Aung}, Han and {Mandelker}, Nir and {Nagai}, Daisuke and {Dekel}, Avishai and {Birnboim}, Yuval},
        title = "{Kelvin-Helmholtz instability in self-gravitating streams}",
      journal = {\mnras},
         year = 2019,
        month = nov,
       volume = {490},
       number = {1},
        pages = {181-201},
          doi = {10.1093/mnras/stz1964},
archivePrefix = {arXiv},
       eprint = {1903.09666},
 primaryClass = {astro-ph.GA},
       adsurl = {https://ui.adsabs.harvard.edu/abs/2019MNRAS.490..181A}
}

@ARTICLE{Towler2024,
       author = {{Towler}, Imogen and {Kay}, Scott T. and {Schaye}, Joop and {Kugel}, Roi and {Schaller}, Matthieu and {Braspenning}, Joey and {Elbers}, Willem and {Frenk}, Carlos S. and {Kwan}, Juliana and {Salcido}, Jaime and {van Daalen}, Marcel P. and {Vandenbroucke}, Bert and {Altamura}, Edoardo},
        title = "{Inferring the dark matter splashback radius from cluster gas and observable profiles in the FLAMINGO simulations}",
      journal = {\mnras},
         year = 2024,
        month = apr,
       volume = {529},
       number = {3},
        pages = {2017-2031},
          doi = {10.1093/mnras/stae654},
archivePrefix = {arXiv},
       eprint = {2312.05126},
 primaryClass = {astro-ph.CO},
       adsurl = {https://ui.adsabs.harvard.edu/abs/2024MNRAS.529.2017T}
}

@ARTICLE{Aung2021_shocksplash,
       author = {{Aung}, Han and {Nagai}, Daisuke and {Lau}, Erwin T.},
        title = "{Shock and splash: gas and dark matter halo boundaries around {\ensuremath{\Lambda}}CDM galaxy clusters}",
      journal = {\mnras},
         year = 2021,
        month = dec,
       volume = {508},
       number = {2},
        pages = {2071-2078},
          doi = {10.1093/mnras/stab2598},
archivePrefix = {arXiv},
       eprint = {2012.00977},
 primaryClass = {astro-ph.CO},
       adsurl = {https://ui.adsabs.harvard.edu/abs/2021MNRAS.508.2071A}
}

@ARTICLE{GalarragaEspinosa2021,
       author = {{Gal{\'a}rraga-Espinosa}, Daniela and {Aghanim}, Nabila and {Langer}, Mathieu and {Tanimura}, Hideki},
        title = "{Properties of gas phases around cosmic filaments at z = 0 in the IllustrisTNG simulation}",
      journal = {\aap},
         year = 2021,
        month = may,
       volume = {649},
          eid = {A117},
        pages = {A117},
          doi = {10.1051/0004-6361/202039781},
archivePrefix = {arXiv},
       eprint = {2010.15139},
 primaryClass = {astro-ph.CO},
       adsurl = {https://ui.adsabs.harvard.edu/abs/2021A&A...649A.117G}
}

@ARTICLE{Tuominen2021,
       author = {{Tuominen}, T. and {Nevalainen}, J. and {Tempel}, E. and {Kuutma}, T. and {Wijers}, N. and {Schaye}, J. and {Hein{\"a}m{\"a}ki}, P. and {Bonamente}, M. and {Ganeshaiah Veena}, P.},
        title = "{An EAGLE view of the missing baryons}",
      journal = {\aap},
         year = 2021,
        month = feb,
       volume = {646},
          eid = {A156},
        pages = {A156},
          doi = {10.1051/0004-6361/202039221},
archivePrefix = {arXiv},
       eprint = {2012.09203},
 primaryClass = {astro-ph.CO},
       adsurl = {https://ui.adsabs.harvard.edu/abs/2021A&A...646A.156T}
}

@ARTICLE{Dave2001_massing_baryons,
       author = {{Dav{\'e}}, Romeel and {Cen}, Renyue and {Ostriker}, Jeremiah P. and
         {Bryan}, Greg L. and {Hernquist}, Lars and {Katz}, Neal and
         {Weinberg}, David H. and {Norman}, Michael L. and {O'Shea}, Brian},
        title = "{Baryons in the Warm-Hot Intergalactic Medium}",
      journal = {\apj},
         year = 2001,
        month = may,
       volume = {552},
       number = {2},
        pages = {473-483},
          doi = {10.1086/320548},
archivePrefix = {arXiv},
       eprint = {astro-ph/0007217},
 primaryClass = {astro-ph},
       adsurl = {https://ui.adsabs.harvard.edu/abs/2001ApJ...552..473D}
}

@ARTICLE{Ostriker1964,
       author = {{Ostriker}, J.},
        title = "{The Equilibrium of Polytropic and Isothermal Cylinders.}",
      journal = {\apj},
         year = 1964,
        month = oct,
       volume = {140},
        pages = {1056},
          doi = {10.1086/148005},
       adsurl = {https://ui.adsabs.harvard.edu/abs/1964ApJ...140.1056O}
}

@ARTICLE{Stodolkiewicz1963,
       author = {{Stod{\'o}lkiewicz}, J.~S.},
        title = "{On the Gravitational Instability of Some Magneto-Hydrodynamical Systems of Astrophysical Interest. Part III.}",
      journal = {\actaa},
         year = 1963,
        month = jan,
       volume = {13},
        pages = {30-54},
       adsurl = {https://ui.adsabs.harvard.edu/abs/1963AcA....13...30S}
}

@ARTICLE{CenOstriker2006,
       author = {{Cen}, Renyue and {Ostriker}, Jeremiah P.},
        title = "{Where Are the Baryons? II. Feedback Effects}",
      journal = {\apj},
         year = 2006,
        month = oct,
       volume = {650},
       number = {2},
        pages = {560-572},
          doi = {10.1086/506505},
archivePrefix = {arXiv},
       eprint = {astro-ph/0601008},
 primaryClass = {astro-ph},
       adsurl = {https://ui.adsabs.harvard.edu/abs/2006ApJ...650..560C}
}

@ARTICLE{GhellerVazza2015,
       author = {{Gheller}, C. and {Vazza}, F. and {Favre}, J. and {Br{\"u}ggen}, M.},
        title = "{Properties of cosmological filaments extracted from Eulerian simulations}",
      journal = {\mnras},
         year = 2015,
        month = oct,
       volume = {453},
       number = {2},
        pages = {1164-1185},
          doi = {10.1093/mnras/stv1646},
archivePrefix = {arXiv},
       eprint = {1507.05806},
 primaryClass = {astro-ph.CO},
       adsurl = {https://ui.adsabs.harvard.edu/abs/2015MNRAS.453.1164G}
}

@ARTICLE{Nicastro2018Nature,
       author = {{Nicastro}, F. and {Kaastra}, J. and {Krongold}, Y. and {Borgani}, S. and
         {Branchini}, E. and {Cen}, R. and {Dadina}, M. and {Danforth}, C.~W. and
         {Elvis}, M. and {Fiore}, F. and {Gupta}, A. and {Mathur}, S. and
         {Mayya}, D. and {Paerels}, F. and {Piro}, L. and {Rosa-Gonzalez}, D. and
         {Schaye}, J. and {Shull}, J.~M. and {Torres-Zafra}, J. and
         {Wijers}, N. and {Zappacosta}, L.},
        title = "{Observations of the missing baryons in the warm-hot intergalactic medium}",
      journal = {\nat},
         year = 2018,
        month = jun,
       volume = {558},
       number = {7710},
        pages = {406-409},
          doi = {10.1038/s41586-018-0204-1},
archivePrefix = {arXiv},
       eprint = {1806.08395},
 primaryClass = {astro-ph.GA},
       adsurl = {https://ui.adsabs.harvard.edu/abs/2018Natur.558..406N}
}

@ARTICLE{White1987,
       author = {{White}, Simon D.~M. and {Frenk}, Carlos S. and {Davis}, Marc and {Efstathiou}, George},
        title = "{Clusters, Filaments, and Voids in a Universe Dominated by Cold Dark Matter}",
      journal = {\apj},
         year = 1987,
        month = feb,
       volume = {313},
        pages = {505},
          doi = {10.1086/164990},
       adsurl = {https://ui.adsabs.harvard.edu/abs/1987ApJ...313..505W}
}

@ARTICLE{GhellerVazza2019_surveyTandNTprops_fils,
       author = {{Gheller}, C. and {Vazza}, F.},
        title = "{A survey of the thermal and non-thermal properties of cosmic filaments}",
      journal = {\mnras},
         year = 2019,
        month = jun,
       volume = {486},
       number = {1},
        pages = {981-1002},
          doi = {10.1093/mnras/stz843},
archivePrefix = {arXiv},
       eprint = {1903.08401},
 primaryClass = {astro-ph.CO},
       adsurl = {https://ui.adsabs.harvard.edu/abs/2019MNRAS.486..981G}
}

@ARTICLE{GalarragaEspinosa2020,
       author = {{Gal{\'a}rraga-Espinosa}, Daniela and {Aghanim}, Nabila and
         {Langer}, Mathieu and {Gouin}, C{\'e}line and {Malavasi}, Nicola},
        title = "{Populations of filaments from the distribution of galaxies in numerical simulations}",
      journal = {\aap},
         year = 2020,
        month = sep,
       volume = {641},
          eid = {A173},
        pages = {A173},
          doi = {10.1051/0004-6361/202037986},
archivePrefix = {arXiv},
       eprint = {2003.09697},
 primaryClass = {astro-ph.CO},
       adsurl = {https://ui.adsabs.harvard.edu/abs/2020A&A...641A.173G}
}

@ARTICLE{Springel2010,
       author = {{Springel}, Volker},
        title = "{E pur si muove: Galilean-invariant cosmological hydrodynamical simulations on a moving mesh}",
      journal = {\mnras},
         year = 2010,
        month = jan,
       volume = {401},
       number = {2},
        pages = {791-851},
          doi = {10.1111/j.1365-2966.2009.15715.x},
archivePrefix = {arXiv},
       eprint = {0901.4107},
 primaryClass = {astro-ph.CO},
       adsurl = {https://ui.adsabs.harvard.edu/abs/2010MNRAS.401..791S}
}

@ARTICLE{Pillepich2018_TNGgalaxies,
       author = {{Pillepich}, Annalisa and {Nelson}, Dylan and {Hernquist}, Lars and {Springel}, Volker and {Pakmor}, R{\"u}diger and {Torrey}, Paul and {Weinberger}, Rainer and {Genel}, Shy and {Naiman}, Jill P. and {Marinacci}, Federico and {Vogelsberger}, Mark},
        title = "{First results from the IllustrisTNG simulations: the stellar mass content of groups and clusters of galaxies}",
      journal = {\mnras},
         year = 2018,
        month = mar,
       volume = {475},
       number = {1},
        pages = {648-675},
          doi = {10.1093/mnras/stx3112},
archivePrefix = {arXiv},
       eprint = {1707.03406},
 primaryClass = {astro-ph.GA},
       adsurl = {https://ui.adsabs.harvard.edu/abs/2018MNRAS.475..648P}
}

@ARTICLE{Feldbrugge2018,
       author = {{Feldbrugge}, Job and {van de Weygaert}, Rien and {Hidding}, Johan and
         {Feldbrugge}, Joost},
        title = "{Caustic Skeleton \&amp; Cosmic Web}",
      journal = {\jcap},
         year = 2018,
        month = may,
       volume = {2018},
       number = {5},
          eid = {027},
        pages = {027},
          doi = {10.1088/1475-7516/2018/05/027},
archivePrefix = {arXiv},
       eprint = {1703.09598},
 primaryClass = {astro-ph.CO},
       adsurl = {https://ui.adsabs.harvard.edu/abs/2018JCAP...05..027F}
}

@ARTICLE{Zakharova2023,
       author = {{Zakharova}, Daria and {Vulcani}, Benedetta and {De Lucia}, Gabriella and {Xie}, Lizhi and {Hirschmann}, Michaela and {Fontanot}, Fabio},
        title = "{The filament determination depends on the tracer: comparing filaments based on dark matter particles and galaxies in the GAEA semi-analytical model}",
      journal = {\mnras},
         year = 2023,
        month = nov,
       volume = {525},
       number = {3},
        pages = {4079-4092},
          doi = {10.1093/mnras/stad2562},
archivePrefix = {arXiv},
       eprint = {2307.05240},
 primaryClass = {astro-ph.GA},
       adsurl = {https://ui.adsabs.harvard.edu/abs/2023MNRAS.525.4079Z}
}

@ARTICLE{Navarro1997_NFW,
       author = {{Navarro}, Julio F. and {Frenk}, Carlos S. and {White}, Simon D.~M.},
        title = "{A Universal Density Profile from Hierarchical Clustering}",
      journal = {\apj},
         year = 1997,
        month = dec,
       volume = {490},
       number = {2},
        pages = {493-508},
          doi = {10.1086/304888},
archivePrefix = {arXiv},
       eprint = {astro-ph/9611107},
 primaryClass = {astro-ph},
       adsurl = {https://ui.adsabs.harvard.edu/abs/1997ApJ...490..493N}
}

@ARTICLE{Zhang2024erosita_fils,
       author = {{Zhang}, X. and {Bulbul}, E. and {Malavasi}, N. and {Ghirardini}, V. and {Comparat}, J. and {Kluge}, M. and {Liu}, A. and {Merloni}, A. and {Zhang}, Y. and {Bahar}, Y.~E. and {Artis}, E. and {Sanders}, J.~S. and {Garrel}, C. and {Balzer}, F. and {Br{\"u}ggen}, M. and {Freyberg}, M. and {Gatuzz}, E. and {Grandis}, S. and {Krippendorf}, S. and {Nandra}, K. and {Ponti}, G. and {Ramos-Ceja}, M. and {Predehl}, P. and {Reiprich}, T.~H. and {Veronica}, A. and {Yeung}, M.~C.~H. and {Zelmer}, S.},
        title = "{The SRG/eROSITA all-sky survey: X-ray emission from the warm-hot phase gas in long cosmic filaments}",
      journal = {\aap},
         year = 2024,
        month = nov,
       volume = {691},
          eid = {A234},
        pages = {A234},
          doi = {10.1051/0004-6361/202450933},
archivePrefix = {arXiv},
       eprint = {2406.00105},
 primaryClass = {astro-ph.HE},
       adsurl = {https://ui.adsabs.harvard.edu/abs/2024A&A...691A.234Z}
}

@ARTICLE{Tanimura2020_Rosat,
       author = {{Tanimura}, H. and {Aghanim}, N. and {Kolodzig}, A. and {Douspis}, M. and {Malavasi}, N.},
        title = "{First detection of stacked X-ray emission from cosmic web filaments}",
      journal = {\aap},
         year = 2020,
        month = nov,
       volume = {643},
          eid = {L2},
        pages = {L2},
          doi = {10.1051/0004-6361/202038521},
archivePrefix = {arXiv},
       eprint = {2011.05343},
 primaryClass = {astro-ph.CO},
       adsurl = {https://ui.adsabs.harvard.edu/abs/2020A&A...643L...2T}
}

@ARTICLE{Planck2015Cosmo,
       author = {{Planck Collaboration} and {Ade}, P.~A.~R. and {Aghanim}, N. and
         {Arnaud}, M. and {Ashdown}, M. and {Aumont}, J. and {Baccigalupi}, C. and
         {Banday}, A.~J. and {Barreiro}, R.~B. and {Bartlett}, J.~G. and
         {Bartolo}, N. and {Battaner}, E. and {Battye}, R. and {Benabed}, K. and
         {Beno{\^\i}t}, A. and {Benoit-L{\'e}vy}, A. and {Bernard}, J. -P. and
         {Bersanelli}, M. and {Bielewicz}, P. and {Bock}, J.~J. and
         {Bonaldi}, A. and {Bonavera}, L. and {Bond}, J.~R. and {Borrill}, J. and
         {Bouchet}, F.~R. and {Boulanger}, F. and {Bucher}, M. and
         {Burigana}, C. and {Butler}, R.~C. and {Calabrese}, E. and
         {Cardoso}, J. -F. and {Catalano}, A. and {Challinor}, A. and
         {Chamballu}, A. and {Chary}, R. -R. and {Chiang}, H.~C. and
         {Chluba}, J. and {Christensen}, P.~R. and {Church}, S. and
         {Clements}, D.~L. and {Colombi}, S. and {Colombo}, L.~P.~L. and
         {Combet}, C. and {Coulais}, A. and {Crill}, B.~P. and {Curto}, A. and
         {Cuttaia}, F. and {Danese}, L. and {Davies}, R.~D. and {Davis}, R.~J. and
         {de Bernardis}, P. and {de Rosa}, A. and {de Zotti}, G. and
         {Delabrouille}, J. and {D{\'e}sert}, F. -X. and {Di Valentino}, E. and
         {Dickinson}, C. and {Diego}, J.~M. and {Dolag}, K. and {Dole}, H. and
         {Donzelli}, S. and {Dor{\'e}}, O. and {Douspis}, M. and {Ducout}, A. and
         {Dunkley}, J. and {Dupac}, X. and {Efstathiou}, G. and {Elsner}, F. and
         {En{\ss}lin}, T.~A. and {Eriksen}, H.~K. and {Farhang}, M. and
         {Fergusson}, J. and {Finelli}, F. and {Forni}, O. and {Frailis}, M. and
         {Fraisse}, A.~A. and {Franceschi}, E. and {Frejsel}, A. and
         {Galeotta}, S. and {Galli}, S. and {Ganga}, K. and {Gauthier}, C. and
         {Gerbino}, M. and {Ghosh}, T. and {Giard}, M. and
         {Giraud-H{\'e}raud}, Y. and {Giusarma}, E. and {Gjerl{\o}w}, E. and
         {Gonz{\'a}lez-Nuevo}, J. and {G{\'o}rski}, K.~M. and {Gratton}, S. and
         {Gregorio}, A. and {Gruppuso}, A. and {Gudmundsson}, J.~E. and
         {Hamann}, J. and {Hansen}, F.~K. and {Hanson}, D. and
         {Harrison}, D.~L. and {Helou}, G. and {Henrot-Versill{\'e}}, S. and
         {Hern{\'a}ndez-Monteagudo}, C. and {Herranz}, D. and {Hildebrand
        t}, S.~R. and {Hivon}, E. and {Hobson}, M. and {Holmes}, W.~A. and
         {Hornstrup}, A. and {Hovest}, W. and {Huang}, Z. and
         {Huffenberger}, K.~M. and {Hurier}, G. and {Jaffe}, A.~H. and
         {Jaffe}, T.~R. and {Jones}, W.~C. and {Juvela}, M. and
         {Keih{\"a}nen}, E. and {Keskitalo}, R. and {Kisner}, T.~S. and
         {Kneissl}, R. and {Knoche}, J. and {Knox}, L. and {Kunz}, M. and
         {Kurki-Suonio}, H. and {Lagache}, G. and {L{\"a}hteenm{\"a}ki}, A. and
         {Lamarre}, J. -M. and {Lasenby}, A. and {Lattanzi}, M. and
         {Lawrence}, C.~R. and {Leahy}, J.~P. and {Leonardi}, R. and
         {Lesgourgues}, J. and {Levrier}, F. and {Lewis}, A. and {Liguori}, M. and
         {Lilje}, P.~B. and {Linden-V{\o}rnle}, M. and {L{\'o}pez-Caniego}, M. and
         {Lubin}, P.~M. and {Mac{\'\i}as-P{\'e}rez}, J.~F. and {Maggio}, G. and
         {Maino}, D. and {Mandolesi}, N. and {Mangilli}, A. and {Marchini}, A. and
         {Maris}, M. and {Martin}, P.~G. and {Martinelli}, M. and
         {Mart{\'\i}nez-Gonz{\'a}lez}, E. and {Masi}, S. and {Matarrese}, S. and
         {McGehee}, P. and {Meinhold}, P.~R. and {Melchiorri}, A. and
         {Melin}, J. -B. and {Mendes}, L. and {Mennella}, A. and
         {Migliaccio}, M. and {Millea}, M. and {Mitra}, S. and
         {Miville-Desch{\^e}nes}, M. -A. and {Moneti}, A. and {Montier}, L. and
         {Morgante}, G. and {Mortlock}, D. and {Moss}, A. and {Munshi}, D. and
         {Murphy}, J.~A. and {Naselsky}, P. and {Nati}, F. and {Natoli}, P. and
         {Netterfield}, C.~B. and {N{\o}rgaard-Nielsen}, H.~U. and
         {Noviello}, F. and {Novikov}, D. and {Novikov}, I. and
         {Oxborrow}, C.~A. and {Paci}, F. and {Pagano}, L. and {Pajot}, F. and
         {Paladini}, R. and {Paoletti}, D. and {Partridge}, B. and {Pasian}, F. and
         {Patanchon}, G. and {Pearson}, T.~J. and {Perdereau}, O. and
         {Perotto}, L. and {Perrotta}, F. and {Pettorino}, V. and
         {Piacentini}, F. and {Piat}, M. and {Pierpaoli}, E. and
         {Pietrobon}, D. and {Plaszczynski}, S. and {Pointecouteau}, E. and
         {Polenta}, G. and {Popa}, L. and {Pratt}, G.~W. and {Pr{\'e}zeau}, G. and
         {Prunet}, S. and {Puget}, J. -L. and {Rachen}, J.~P. and
         {Reach}, W.~T. and {Rebolo}, R. and {Reinecke}, M. and
         {Remazeilles}, M. and {Renault}, C. and {Renzi}, A. and
         {Ristorcelli}, I. and {Rocha}, G. and {Rosset}, C. and {Rossetti}, M. and
         {Roudier}, G. and {Rouill{\'e} d'Orfeuil}, B. and {Rowan-Robinson}, M. and
         {Rubi{\~n}o-Mart{\'\i}n}, J.~A. and {Rusholme}, B. and {Said}, N. and
         {Salvatelli}, V. and {Salvati}, L. and {Sandri}, M. and {Santos}, D. and
         {Savelainen}, M. and {Savini}, G. and {Scott}, D. and
         {Seiffert}, M.~D. and {Serra}, P. and {Shellard}, E.~P.~S. and
         {Spencer}, L.~D. and {Spinelli}, M. and {Stolyarov}, V. and
         {Stompor}, R. and {Sudiwala}, R. and {Sunyaev}, R. and {Sutton}, D. and
         {Suur-Uski}, A. -S. and {Sygnet}, J. -F. and {Tauber}, J.~A. and
         {Terenzi}, L. and {Toffolatti}, L. and {Tomasi}, M. and {Tristram}, M. and
         {Trombetti}, T. and {Tucci}, M. and {Tuovinen}, J. and
         {T{\"u}rler}, M. and {Umana}, G. and {Valenziano}, L. and
         {Valiviita}, J. and {Van Tent}, F. and {Vielva}, P. and {Villa}, F. and
         {Wade}, L.~A. and {Wandelt}, B.~D. and {Wehus}, I.~K. and {White}, M. and
         {White}, S.~D.~M. and {Wilkinson}, A. and {Yvon}, D. and {Zacchei}, A. and
         {Zonca}, A.},
        title = "{Planck 2015 results. XIII. Cosmological parameters}",
      journal = {\aap},
         year = "2016",
        month = "Sep",
       volume = {594},
          eid = {A13},
        pages = {A13},
          doi = {10.1051/0004-6361/201525830},
archivePrefix = {arXiv},
       eprint = {1502.01589},
 primaryClass = {astro-ph.CO},
       adsurl = {https://ui.adsabs.harvard.edu/abs/2016A&A...594A..13P}
}

@ARTICLE{Malavasi2020_sdss,
       author = {{Malavasi}, Nicola and {Aghanim}, Nabila and {Douspis}, Marian and
         {Tanimura}, Hideki and {Bonjean}, Victor},
        title = "{Characterising filaments in the SDSS volume from the galaxy distribution}",
      journal = {\aap},
         year = 2020,
        month = oct,
       volume = {642},
          eid = {A19},
        pages = {A19},
          doi = {10.1051/0004-6361/202037647},
archivePrefix = {arXiv},
       eprint = {2002.01486},
 primaryClass = {astro-ph.CO},
       adsurl = {https://ui.adsabs.harvard.edu/abs/2020A&A...642A..19M}
}

@ARTICLE{Aghanim2024_shapley,
       author = {{Aghanim}, N. and {Tuominen}, T. and {Bonjean}, V. and {Gouin}, C. and {Bonnaire}, T. and {Einasto}, M.},
        title = "{Dissecting a miniature universe: A multi-wavelength view of galaxy quenching in the Shapley supercluster}",
      journal = {\aap},
         year = 2024,
        month = sep,
       volume = {689},
          eid = {A332},
        pages = {A332},
          doi = {10.1051/0004-6361/202348672},
archivePrefix = {arXiv},
       eprint = {2402.18455},
 primaryClass = {astro-ph.CO},
       adsurl = {https://ui.adsabs.harvard.edu/abs/2024A&A...689A.332A}
}

@ARTICLE{Lebeau2024_sp,
       author = {{Lebeau}, Th{\'e}o and {Ettori}, Stefano and {Aghanim}, Nabila and {Sorce}, Jenny G.},
        title = "{Can the splashback radius be an observable boundary of galaxy clusters?}",
      journal = {\aap},
         year = 2024,
        month = sep,
       volume = {689},
          eid = {A19},
        pages = {A19},
          doi = {10.1051/0004-6361/202450146},
archivePrefix = {arXiv},
       eprint = {2403.18648},
 primaryClass = {astro-ph.CO},
       adsurl = {https://ui.adsabs.harvard.edu/abs/2024A&A...689A..19L}
}

@ARTICLE{Migkas2025,
       author = {{Migkas}, K. and {Pacaud}, F. and {Tuominen}, T. and {Aghanim}, N.},
        title = "{Detection of pure warm-hot intergalactic medium emission from a 7.2 Mpc long filament in the Shapley supercluster using X-ray spectroscopy}",
      journal = {\aap},
         year = 2025,
        month = jun,
       volume = {698},
          eid = {A270},
        pages = {A270},
          doi = {10.1051/0004-6361/202554944},
archivePrefix = {arXiv},
       eprint = {2506.14917},
 primaryClass = {astro-ph.CO},
       adsurl = {https://ui.adsabs.harvard.edu/abs/2025A&A...698A.270M}
}

@ARTICLE{Wang2026_whim,
       author = {{Wang}, Lile and {Cen}, Renyue},
        title = "{High Resolution Grid-based Simulations of the Warm-Hot Intergalactic Medium}",
      journal = {arXiv e-prints},
         year = 2026,
        month = sep,
          eid = {arXiv:2609.21784},
        pages = {arXiv:2609.21784},
          doi = {10.48550/arXiv.2609.21784},
archivePrefix = {arXiv},
       eprint = {2609.21784},
 primaryClass = {astro-ph.CO},
       adsurl = {https://ui.adsabs.harvard.edu/abs/2026arXiv260921784W}
}

@ARTICLE{Nagai2006,
       author = {{Nagai}, Daisuke},
        title = "{The Impact of Galaxy Formation on the Sunyaev-Zel'dovich Effect of Galaxy Clusters}",
      journal = {\apj},
         year = 2006,
        month = oct,
       volume = {650},
       number = {2},
        pages = {538-549},
          doi = {10.1086/506467},
archivePrefix = {arXiv},
       eprint = {astro-ph/0512208},
 primaryClass = {astro-ph},
       adsurl = {https://ui.adsabs.harvard.edu/abs/2006ApJ...650..538N}
}

@ARTICLE{DeGraaff2019,
       author = {{de Graaff}, Anna and {Cai}, Yan-Chuan and {Heymans}, Catherine and {Peacock}, John A.},
        title = "{Probing the missing baryons with the Sunyaev-Zel'dovich effect from filaments}",
      journal = {\aap},
         year = 2019,
        month = apr,
       volume = {624},
          eid = {A48},
        pages = {A48},
          doi = {10.1051/0004-6361/201935159},
archivePrefix = {arXiv},
       eprint = {1709.10378},
 primaryClass = {astro-ph.CO},
       adsurl = {https://ui.adsabs.harvard.edu/abs/2019A&A...624A..48D}
}

@ARTICLE{Tanimura2019,
       author = {{Tanimura}, Hideki and {Hinshaw}, Gary and {McCarthy}, Ian G. and
         {Van Waerbeke}, Ludovic and {Aghanim}, Nabila and {Ma}, Yin-Zhe and
         {Mead}, Alexander and {Hojjati}, Alireza and {Tr{\"o}ster}, Tilman},
        title = "{A search for warm/hot gas filaments between pairs of SDSS Luminous Red Galaxies}",
      journal = {\mnras},
         year = "2019",
        month = "Feb",
       volume = {483},
       number = {1},
        pages = {223-234},
          doi = {10.1093/mnras/sty3118},
archivePrefix = {arXiv},
       eprint = {1709.05024},
 primaryClass = {astro-ph.CO},
       adsurl = {https://ui.adsabs.harvard.edu/abs/2019MNRAS.483..223T}
}

@ARTICLE{Tanimura2022_eRosita,
       author = {{Tanimura}, H. and {Aghanim}, N. and {Douspis}, M. and {Malavasi}, N.},
        title = "{X-ray emission from cosmic web filaments in SRG/eROSITA data}",
      journal = {\aap},
         year = 2022,
        month = nov,
       volume = {667},
          eid = {A161},
        pages = {A161},
          doi = {10.1051/0004-6361/202244158},
archivePrefix = {arXiv},
       eprint = {2206.00084},
 primaryClass = {astro-ph.CO},
       adsurl = {https://ui.adsabs.harvard.edu/abs/2022A&A...667A.161T}
}

@ARTICLE{GhellerVazza2016,
       author = {{Gheller}, C. and {Vazza}, F. and {Br{\"u}ggen}, M. and {Alpaslan}, M. and
         {Holwerda}, B.~W. and {Hopkins}, A.~M. and {Liske}, J.},
        title = "{Evolution of cosmic filaments and of their galaxy population from MHD cosmological simulations}",
      journal = {\mnras},
         year = "2016",
        month = "Oct",
       volume = {462},
       number = {1},
        pages = {448-463},
          doi = {10.1093/mnras/stw1595},
archivePrefix = {arXiv},
       eprint = {1607.01406},
 primaryClass = {astro-ph.CO},
       adsurl = {https://ui.adsabs.harvard.edu/abs/2016MNRAS.462..448G}
}

@ARTICLE{Tempel2016bisous,
       author = {{Tempel}, E. and {Stoica}, R.~S. and {Kipper}, R. and {Saar}, E.},
        title = "{Bisous model-Detecting filamentary patterns in point processes}",
      journal = {Astronomy and Computing},
         year = "2016",
        month = "Jul",
       volume = {16},
        pages = {17-25},
          doi = {10.1016/j.ascom.2016.03.004},
archivePrefix = {arXiv},
       eprint = {1603.08957},
 primaryClass = {astro-ph.CO},
       adsurl = {https://ui.adsabs.harvard.edu/abs/2016A&C....16...17T}
}

@ARTICLE{Pratt2010,
       author = {{Pratt}, G.~W. and {Arnaud}, M. and {Piffaretti}, R. and {B{\"o}hringer}, H. and {Ponman}, T.~J. and {Croston}, J.~H. and {Voit}, G.~M. and {Borgani}, S. and {Bower}, R.~G.},
        title = "{Gas entropy in a representative sample of nearby X-ray galaxy clusters (REXCESS): relationship to gas mass fraction}",
      journal = {\aap},
         year = 2010,
        month = feb,
       volume = {511},
          eid = {A85},
        pages = {A85},
          doi = {10.1051/0004-6361/200913309},
archivePrefix = {arXiv},
       eprint = {0909.3776},
 primaryClass = {astro-ph.CO},
       adsurl = {https://ui.adsabs.harvard.edu/abs/2010A&A...511A..85P}
}

@ARTICLE{Zhu2021_entropy_clusters,
       author = {{Zhu}, Zhenghao and {Xu}, Haiguang and {Hu}, Dan and {Shan}, Chenxi and {Zhu}, Yongkai and {Fan}, Shida and {Zhao}, Yuanyuan and {Gu}, Liyi and {Wu}, Xiang-Ping},
        title = "{A Study of Gas Entropy Profiles of 47 Galaxy Clusters and Groups out to the Virial Radius}",
      journal = {\apj},
         year = 2021,
        month = feb,
       volume = {908},
       number = {1},
          eid = {17},
        pages = {17},
          doi = {10.3847/1538-4357/abd327},
archivePrefix = {arXiv},
       eprint = {2101.05947},
 primaryClass = {astro-ph.CO},
       adsurl = {https://ui.adsabs.harvard.edu/abs/2021ApJ...908...17Z}
}

@ARTICLE{Sen2026_1halo2boundaries,
       author = {{Sen}, Siddhant and {Adhikari}, Susmita and {Nagai}, Daisuke and {Diemer}, Benedikt},
        title = "{One Halo, Two Boundaries: Relating Accretion Shocks and Splashback Radii in Galaxy Clusters}",
      journal = {arXiv e-prints},
         year = 2026,
        month = feb,
          eid = {arXiv:2602.18609},
        pages = {arXiv:2602.18609},
          doi = {10.48550/arXiv.2602.18609},
archivePrefix = {arXiv},
       eprint = {2602.18609},
 primaryClass = {astro-ph.CO},
       adsurl = {https://ui.adsabs.harvard.edu/abs/2026arXiv260218609S}
}

@ARTICLE{Voit2005_entropy,
       author = {{Voit}, G. Mark and {Kay}, Scott T. and {Bryan}, Greg L.},
        title = "{The baseline intracluster entropy profile from gravitational structure formation}",
      journal = {\mnras},
         year = 2005,
        month = dec,
       volume = {364},
       number = {3},
        pages = {909-916},
          doi = {10.1111/j.1365-2966.2005.09621.x},
archivePrefix = {arXiv},
       eprint = {astro-ph/0511252},
 primaryClass = {astro-ph},
       adsurl = {https://ui.adsabs.harvard.edu/abs/2005MNRAS.364..909V}
}

@ARTICLE{YangQirui2025_width_filaments,
       author = {{Yang}, Qi-Rui and {Zhu}, Weishan and {Yu}, Guang-Yao and {Mo}, Jian-Feng and {Zheng}, Yi and {Feng}, Long-Long},
        title = "{On the Width and Profiles of Cosmic Filaments}",
      journal = {\apj},
         year = 2025,
        month = aug,
       volume = {989},
       number = {2},
          eid = {187},
        pages = {187},
          doi = {10.3847/1538-4357/adeca3},
archivePrefix = {arXiv},
       eprint = {2507.02476},
 primaryClass = {astro-ph.CO},
       adsurl = {https://ui.adsabs.harvard.edu/abs/2025ApJ...989..187Y}
}

@article{Bond1996,
       author = {{Bond}, J. Richard and {Kofman}, Lev and {Pogosyan}, Dmitry},
        title = "{How filaments of galaxies are woven into the cosmic web}",
      journal = {\nat},
         year = "1996",
        month = "Apr",
       volume = {380},
       number = {6575},
        pages = {603-606},
          doi = {10.1038/380603a0},
archivePrefix = {arXiv},
       eprint = {astro-ph/9512141},
 primaryClass = {astro-ph},
       adsurl = {https://ui.adsabs.harvard.edu/abs/1996Natur.380..603B}
}

@ARTICLE{Lu2024_fils,
       author = {{Lu}, Yue Samuel and {Mandelker}, Nir and {Oh}, Siang Peng and {Dekel}, Avishai and {van den Bosch}, Frank C. and {Springel}, Volker and {Nagai}, Daisuke and {van de Voort}, Freeke},
        title = "{The structure and dynamics of massive high-z cosmic-web filaments: three radial zones in filament cross-sections}",
      journal = {\mnras},
         year = 2024,
        month = feb,
       volume = {527},
       number = {4},
        pages = {11256-11287},
          doi = {10.1093/mnras/stad3779},
archivePrefix = {arXiv},
       eprint = {2306.03966},
 primaryClass = {astro-ph.CO},
       adsurl = {https://ui.adsabs.harvard.edu/abs/2024MNRAS.52711256L}
}

@article{Colberg2005,
       author = {{Colberg}, J{\"o}rg M. and {Krughoff}, K. Simon and
         {Connolly}, Andrew J.},
        title = "{Intercluster filaments in a {\ensuremath{\Lambda}}CDM Universe}",
      journal = {\mnras},
         year = "2005",
        month = "May",
       volume = {359},
       number = {1},
        pages = {272-282},
          doi = {10.1111/j.1365-2966.2005.08897.x},
archivePrefix = {arXiv},
       eprint = {astro-ph/0406665},
 primaryClass = {astro-ph},
       adsurl = {https://ui.adsabs.harvard.edu/abs/2005MNRAS.359..272C}
}

@article{Bonjean2018,
       author = {{Bonjean}, V. and {Aghanim}, N. and {Salom{\'e}}, P. and {Douspis}, M. and
         {Beelen}, A.},
        title = "{Gas and galaxies in filaments between clusters of galaxies. The study of A399-A401}",
      journal = {\aap},
         year = "2018",
        month = "Jan",
       volume = {609},
          eid = {A49},
        pages = {A49},
          doi = {10.1051/0004-6361/201731699},
archivePrefix = {arXiv},
       eprint = {1710.08699},
 primaryClass = {astro-ph.CO},
       adsurl = {https://ui.adsabs.harvard.edu/abs/2018A&A...609A..49B}
}

@ARTICLE{Zeldovich1970,
       author = {{Zel'dovich}, Ya. B.},
        title = "{Gravitational instability: An approximate theory for large density perturbations.}",
      journal = {\aap},
         year = 1970,
        month = mar,
       volume = {5},
        pages = {84-89},
       adsurl = {https://ui.adsabs.harvard.edu/abs/1970A&A.....5...84Z}
}

@article{Cautun2014,
       author = {{Cautun}, Marius and {van de Weygaert}, Rien and {Jones}, Bernard J.~T. and
         {Frenk}, Carlos S.},
        title = "{Evolution of the cosmic web}",
      journal = {\mnras},
         year = "2014",
        month = "Jul",
       volume = {441},
       number = {4},
        pages = {2923-2973},
          doi = {10.1093/mnras/stu768},
archivePrefix = {arXiv},
       eprint = {1401.7866},
 primaryClass = {astro-ph.CO},
       adsurl = {https://ui.adsabs.harvard.edu/abs/2014MNRAS.441.2923C}
      }

@article{Nelson2018,
       author = {{Nelson}, Dylan and {Pillepich}, Annalisa and {Springel}, Volker and
         {Weinberger}, Rainer and {Hernquist}, Lars and {Pakmor}, R{\"u}diger and
         {Genel}, Shy and {Torrey}, Paul and {Vogelsberger}, Mark and
         {Kauffmann}, Guinevere and {Marinacci}, Federico and {Naiman}, Jill},
        title = "{First results from the IllustrisTNG simulations: the galaxy colour bimodality}",
      journal = {\mnras},
         year = "2018",
        month = "Mar",
       volume = {475},
       number = {1},
        pages = {624-647},
          doi = {10.1093/mnras/stx3040},
archivePrefix = {arXiv},
       eprint = {1707.03395},
 primaryClass = {astro-ph.GA},
       adsurl = {https://ui.adsabs.harvard.edu/abs/2018MNRAS.475..624N}
}

@ARTICLE{Feldbrugge2025,
       author = {{Feldbrugge}, Job and {van de Weygaert}, Rien},
        title = "{What makes a cosmic filament? The dynamical origin and identity of filaments ─ I. Fundamentals in 2D}",
      journal = {\mnras},
         year = 2025,
        month = may,
       volume = {539},
       number = {2},
        pages = {873-901},
          doi = {10.1093/mnras/staf536},
archivePrefix = {arXiv},
       eprint = {2405.20475},
 primaryClass = {astro-ph.CO},
       adsurl = {https://ui.adsabs.harvard.edu/abs/2025MNRAS.539..873F}
}

@ARTICLE{Neyrinck2012_Origami,
       author = {{Neyrinck}, Mark C.},
        title = "{Origami constraints on the initial-conditions arrangement of dark-matter caustics and streams}",
      journal = {\mnras},
         year = 2012,
        month = nov,
       volume = {427},
       number = {1},
        pages = {494-501},
          doi = {10.1111/j.1365-2966.2012.21956.x},
archivePrefix = {arXiv},
       eprint = {1202.3364},
 primaryClass = {astro-ph.CO},
       adsurl = {https://ui.adsabs.harvard.edu/abs/2012MNRAS.427..494N}
}

@ARTICLE{Tinker2008,
       author = {{Tinker}, Jeremy and {Kravtsov}, Andrey V. and {Klypin}, Anatoly and {Abazajian}, Kevork and {Warren}, Michael and {Yepes}, Gustavo and {Gottl{\"o}ber}, Stefan and {Holz}, Daniel E.},
        title = "{Toward a Halo Mass Function for Precision Cosmology: The Limits of Universality}",
      journal = {\apj},
         year = 2008,
        month = dec,
       volume = {688},
       number = {2},
        pages = {709-728},
          doi = {10.1086/591439},
archivePrefix = {arXiv},
       eprint = {0803.2706},
 primaryClass = {astro-ph},
       adsurl = {https://ui.adsabs.harvard.edu/abs/2008ApJ...688..709T}
}

@article{Sousbie2011,
       author = {{Sousbie}, T.},
        title = "{The persistent cosmic web and its filamentary structure - I. Theory and implementation}",
      journal = {\mnras},
         year = "2011",
        month = "Jun",
       volume = {414},
       number = {1},
        pages = {350-383},
          doi = {10.1111/j.1365-2966.2011.18394.x},
archivePrefix = {arXiv},
       eprint = {1009.4015},
 primaryClass = {astro-ph.CO},
       adsurl = {https://ui.adsabs.harvard.edu/abs/2011MNRAS.414..350S}
}

@article{Sousbie2011b,
       author = {{Sousbie}, T. and {Pichon}, C. and {Kawahara}, H.},
        title = "{The persistent cosmic web and its filamentary structure - II. Illustrations}",
      journal = {\mnras},
         year = 2011,
        month = Jun,
       volume = {414},
        pages = {384-403},
          doi = {10.1111/j.1365-2966.2011.18395.x},
archivePrefix = {arXiv},
       eprint = {1009.4014},
 primaryClass = {astro-ph.CO},
       adsurl = {https://ui.adsabs.harvard.edu/\#abs/2011MNRAS.414..384S}
}

@ARTICLE{Nelson2019_TNGdata_release,
       author = {{Nelson}, Dylan and {Springel}, Volker and {Pillepich}, Annalisa and
         {Rodriguez-Gomez}, Vicente and {Torrey}, Paul and {Genel}, Shy and
         {Vogelsberger}, Mark and {Pakmor}, Ruediger and {Marinacci}, Federico and
         {Weinberger}, Rainer and {Kelley}, Luke and {Lovell}, Mark and
         {Diemer}, Benedikt and {Hernquist}, Lars},
        title = "{The IllustrisTNG simulations: public data release}",
      journal = {Computational Astrophysics and Cosmology},
         year = "2019",
        month = "May",
       volume = {6},
       number = {1},
          eid = {2},
        pages = {2},
          doi = {10.1186/s40668-019-0028-x},
archivePrefix = {arXiv},
       eprint = {1812.05609},
 primaryClass = {astro-ph.GA},
       adsurl = {https://ui.adsabs.harvard.edu/abs/2019ComAC...6....2N}
}

@ARTICLE{Xu2026_filamentsDMpro,
       author = {{Xu}, Peng and {Jiang}, Fangzhou and {Hasan}, Farhanul and {Woo}, Joanna and {Hellinger}, Douglas and {Primack}, Joel R. and {Faber}, Sandra M. and {Koo}, David},
        title = "{Universal Dark-matter Density Profiles of Cosmic Filaments}",
      journal = {arXiv e-prints},
         year = 2026,
        month = apr,
          eid = {arXiv:2604.05033},
        pages = {arXiv:2604.05033},
          doi = {10.48550/arXiv.2604.05033},
archivePrefix = {arXiv},
       eprint = {2604.05033},
 primaryClass = {astro-ph.CO},
       adsurl = {https://ui.adsabs.harvard.edu/abs/2026arXiv260405033X}
}

@ARTICLE{Virtanen2020SciPy,
  author  = {Virtanen, Pauli and Gommers, Ralf and Oliphant, Travis E. and
            Haberland, Matt and Reddy, Tyler and Cournapeau, David and
            Burovski, Evgeni and Peterson, Pearu and Weckesser, Warren and
            Bright, Jonathan and {van der Walt}, St{\'e}fan J. and
            Brett, Matthew and Wilson, Joshua and Millman, K. Jarrod and
            Mayorov, Nikolay and Nelson, Andrew R. J. and Jones, Eric and
            Kern, Robert and Larson, Eric and Carey, C J and
            Polat, {\.I}lhan and Feng, Yu and Moore, Eric W. and
            {VanderPlas}, Jake and Laxalde, Denis and Perktold, Josef and
            Cimrman, Robert and Henriksen, Ian and Quintero, E. A. and
            Harris, Charles R. and Archibald, Anne M. and
            Ribeiro, Ant{\^o}nio H. and Pedregosa, Fabian and
            {van Mulbregt}, Paul and {SciPy 1.0 Contributors}},
  title   = {{{SciPy} 1.0: Fundamental Algorithms for Scientific
            Computing in Python}},
  journal = {Nature Methods},
  year    = {2020},
  volume  = {17},
  pages   = {261--272},
  adsurl  = {https://rdcu.be/b08Wh},
  doi     = {10.1038/s41592-019-0686-2},
}
\FloatBarrier

\begin{appendix}
\section{Fit results of filament line-mass function}\label{APP:MCMC}

In Sect.~\ref{SubSect:LineMassFunction} we fit different functional forms to the gaseous line-mass function of Fig.~\ref{Fig:Lambda_fils}. Here we provide the explicit functional forms and best fit parameter values. 
We fit a simple Schechter function (Eq.~\ref{Eq:schechter}), the generalised Schechter function (Eq.~\ref{Eq:gen_schechter}), the Tinker-like model (Eq.~\ref{Eq:Tinker}), a double power-law (Eq.~\ref{Eq:douple_pl}), and the GNFW functional form (Eq.~\ref{Eq:GNFW}). The best fit parameter values are reported in Table~\ref{Tab:fit_mass}.

\begin{equation}\label{Eq:schechter}
    \frac{dn}{d\Lambda}= C \left(\frac{\Lambda}{\Lambda_c}\right)^\alpha \exp\!\left(-\frac{\Lambda}{\Lambda_c}\right).
\end{equation}

\begin{equation}\label{Eq:gen_schechter}
    \frac{dn}{d\Lambda}=C \left(\frac{\Lambda}{\Lambda_c}\right)^\alpha \exp\!\left[-\left(\frac{\Lambda}{\Lambda_c}\right)^\beta\right].
\end{equation}

\begin{equation}\label{Eq:douple_pl}
    \frac{dn}{d\Lambda}=C \left[\left(\frac{\Lambda}{\Lambda_c}\right)^\alpha+\left(\frac{\Lambda}{\Lambda_c}\right)^\beta\right]^{-1}.
\end{equation}

\begin{equation}\label{Eq:GNFW}
    \frac{dn}{d\Lambda}= \frac{C}{ \left(\frac{\Lambda}{\Lambda_c}\right)^\alpha \left[ 1 + \left( \frac{\Lambda}{\Lambda_c} \right)^\gamma  \right]^{\frac{\beta-\alpha}{\gamma}} }.
\end{equation}

\begin{table*}
	\centering
	\caption{Gaseous line-mass function fit results. We report the best-fitting coefficients, with the corresponding $1\sigma$ errors.}
	\label{Tab:fit_mass}
	\begin{tabular}{lccccc}
		\hline
		Model & C & $\Lambda_c$ & $\alpha$ & $\beta$ & $\gamma$ \\
		\hline
		Schechter (Eq.~\ref{Eq:schechter}) & $3.80_{-0.20}^{+0.20} \times 10^{-16}$ & $5.89_{-0.22}^{+0.23} \times 10^{11}$ & $0.013_{-0.038}^{+0.039}$ & - & - \\
        Gen. Schechter (Eq.~\ref{Eq:gen_schechter}) & $2.78_{-1.55}^{+2.20} \times 10^{-16}$ & $1.53_{-0.77}^{+1.50} \times 10^{10}$ & $1.26_{-0.21}^{+0.21}$ & $0.455_{-0.035}^{+0.043}$ \\
        Tinker (Eq.~\ref{Eq:Tinker}) & $8.23_{-4.45}^{+4.91} \times 10^{-17}$ & $5.84_{-2.40}^{+2.54} \times 10^{9}$ & $1.55_{-0.10}^{+0.14}$ & $0.409_{-0.021}^{+0.017}$ & - \\
        Double PL (Eq.~\ref{Eq:douple_pl}) & $1.84_{-0.12}^{+0.13} \times 10^{-16}$ & $8.57_{-0.36}^{+0.37} \times 10^{11}$ & $0.240_{-0.040}^{+0.038}$ & $3.28_{-0.07}^{+0.07}$ & - \\
        GNFW (Eq.~\ref{Eq:GNFW}) & $8.22_{-1.94}^{+1.26} \times 10^{-16}$ & $1.26_{-0.17}^{+0.25} \times 10^{12}$ & $-0.332_{-0.061}^{+0.082}$ & $4.82_{-0.40}^{+0.52}$ & $1.20_{-0.07}^{+0.11}$ \\
		\hline
	\end{tabular}
\end{table*}

\section{Galaxy number-density profile by bins of $\Lambda$}\label{APP:ngal}

\begin{figure}
    \centering
    \includegraphics[width=0.5\linewidth]{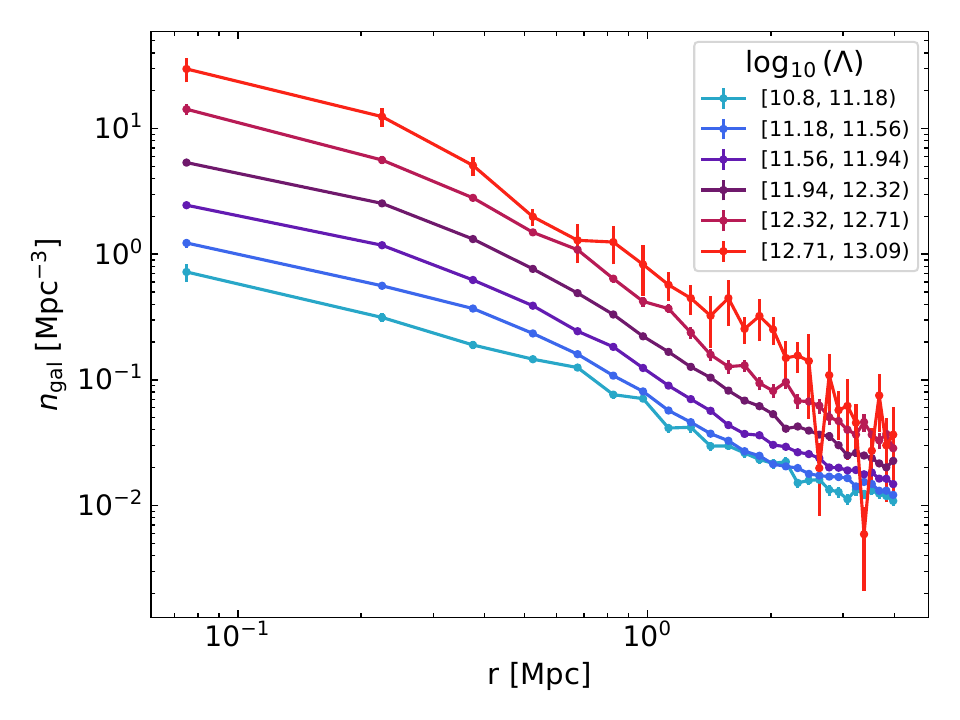}
    \caption{Galaxy number-density profile by bins of $\Lambda$.}
    \label{FigAPP:ngal}
\end{figure}

The profiles are shown in Fig.~\ref{FigAPP:ngal}.
For the computation of the galaxy density profiles, we use all galaxies with stellar masses above $10^9\,\Msun{}$ and we follow the method introduced in \cite{GalarragaEspinosa2020}.


\end{appendix}

\FloatBarrier

\end{document}